\documentclass[aps,prb,reprint,superscriptaddress]{revtex4-1}

\usepackage[utf8]{inputenc}
\usepackage[usenames, dvipsnames, svgnames, table]{xcolor}
\usepackage{mathtools}
\usepackage{amsfonts}
\usepackage{graphicx}
\usepackage[colorlinks=true, citecolor=Blue,
            linkcolor=BrickRed, urlcolor=ForestGreen]{hyperref}
\usepackage{siunitx}

\usepackage[shortlabels]{enumitem}

\DeclareSIUnit\ppm{ppm}

\begin{document}

\footnotetext[5]{An excellent primer to state-space representations of dynamical systems can be found in~\cite{Bechhoefer2005}.}


\title{Direct approach to realising quantum filters for high-precision measurements}


\newcommand{\uob}{Institute for Gravitational Wave Astronomy, School of Physics and
Astronomy, University of Birmingham, Birmingham B15 2TT, United Kingdom}
\author{Joe Bentley}
\affiliation{\uob}
\email[Corresponding author: ]{jbentley@star.sr.bham.ac.uk}
\author{Hendra Nurdin}
\affiliation{School of Electrical Engineering and Telecommunications, University of New South Wales, Sydney 2052, Australia}
\author{Yanbei Chen}
\affiliation{Theoretical Astrophysics 350-17, California Institute of Technology, Pasadena, California 91125, USA}
\author{Haixing Miao}
\affiliation{\uob}


\date{\today}

\begin{abstract}
Quantum noise sets
a fundamental limit to the sensitivity of high-precision measurements. Suppressing it can be achieved by using non-classical states and quantum filters, which 
modify both the noise and signal response. We find an approach to realising quantum filters directly
from their frequency-domain transfer functions, utilising techniques developed by the quantum control community. It not only allows us to construct quantum filters that defy intuition, but also opens a path towards the systematic design of optimal quantum measurement devices. As an illustration, we show a new optical realisation of an active unstable filter with anomalous dispersion, proposed for improving the quantum-limited sensitivity of gravitational-wave detectors. 
\end{abstract}


\maketitle


\section{Introduction}

\begin{figure}[!b]
    \centering
    \includegraphics[width=\linewidth]{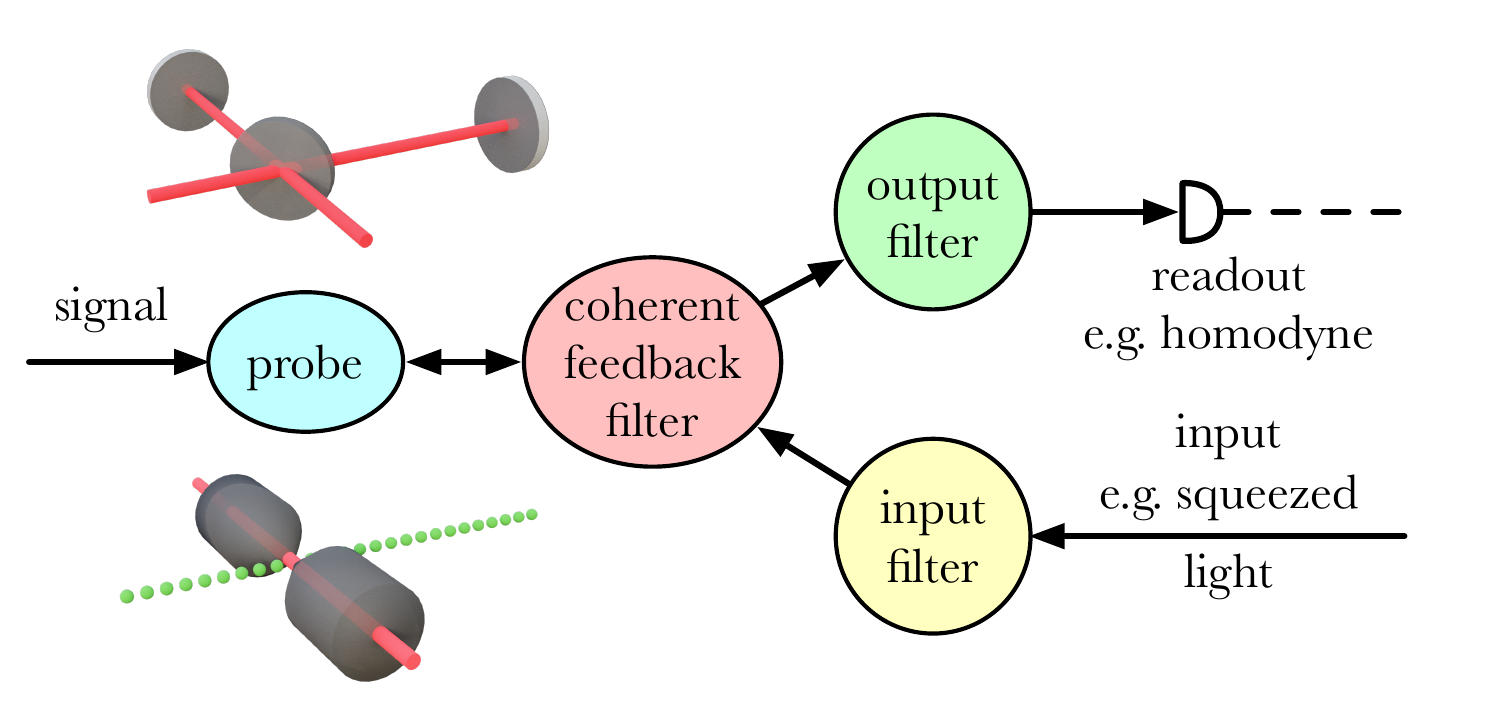}
    \caption{Flowchart illustrating the different places quantum filters can be used within a quantum measurement device. We consider a generic device consisting of a probe (e.g.\, a mirror-endowed test mass or an atomic ensemble) coupled to some classical signal, which receives an input (e.g.~non-classical squeezed light) and whose output field is measured by the readout scheme (e.g.homodyne readout).}
    \label{fig:quantum-filters}
\end{figure}

In high-precision measurements, our understanding of physics is predominantly limited by quantum noise, arising due to the fundamental quantum fluctuations of the probing fields~\cite{Braginsky, Caves1980, Gardiner1991, Clerk2010a}. This is particularly true for laser interferometric gravitational-wave detectors~\cite{Adhikari2014} where the quantum shot noise dominates at high frequencies due to the positive dispersion of the arm cavities~\cite{Miao}. Quantum and classical noises are also limiting factors in quantum optomechanical  experiments~\cite{ Chen2013, Aspelmeyer2014} and searches for new physics using an interferometer~\cite{Sikivie1983, Derocco2018}. To achieve a maximal signal-to-noise ratio, it is essential to engineer the frequency-dependent response of the measurement devices depending on the frequency content of the signal being measured. For example, advanced gravitational wave detectors are tuned to have maximum sensitivity in a frequency range containing the binary black hole inspiral waveform, however not all of the binary neutron star inspiral waveform is observed~\cite{TheLIGOScientificCollaboration2019}. Quantum filters are designed to engineer this response. As illustrated in Fig.~\ref{fig:quantum-filters}, there are three ways that the measurement device can be augmented with quantum filters. First, the input filter, coupling the noise input to the probe degrees of freedom, shapes how the quantum fluctuations enter the device. Next, the coherent feedback filter, coupled to the probe degrees of freedom and input-output fields, modifies the dynamics of the probe~\cite{James2008,Mabuchi2008a,Hamerly2012,Jacobs2014}. This can enhance the response to the signal of interest when the quantum system is converted into a probe coupled to a classical signal. Finally, the output filter, coupling the probe degrees of freedom to the readout port, modifies the response of readout to the detector's the output field. As a simple example, an optical Fabry-Perot cavity is used as an 
input filter to in implementing frequency-dependent squeezed light~\cite{Kimble2000,Oelker2016c,Schnabel2017}.

Until now, formulating a physical realisation of a given quantum filter with a desired frequency response required a combination of intuition and prior experience, making more complicated frequency responses difficult to engineer. We adopt the general formalism for describing linear stochastic quantum networks and the synthesis of such networks, recently developed by the quantum control community~\cite{Gough2007, James2008, Gough2009, Gough2009a, Tezak2012a, Combes2016, Nurdin2009, Nurdin2010a,Nurdin2010b,Nurdin2016, Grivopoulos2016,Nurdin2017,Grivopoulos2017,Petersen2018}. This allows us in this paper to develop a formalism for systematically realising quantum filters for high-precision measurements directly from their frequency-domain transfer matrices. Although all of the components of the formalism existed previously this is the first time the entire process of going from the transfer matrix to physical realisation has been written down, and the concept of using the formalism to produce quantum filters for high-precision measurement is a totally original contribution, as well as the method of transforming to an unrealisable state-space to a realisable one.

The starting point of this approach is to convert the frequency-domain transfer matrix into a
state-space representation. However, such a mapping is not unique, and in fact infinitely many different state-space representations exist.  
We apply the concept of physical realisability, which was first introduced in~\cite{James2008} and further discussed in~\cite{Gough2009}\cite[Chapter 2]{Nurdin2017}. It tells us whether a given time-domain state-space representation of the system obeys quantum mechanics.
Requiring that the state-space is physically realisable sets a significant constraint on the range of this mapping. However, when a state-space realisation satisfies certain conditions, following \cite{Shaiju2012}, we can transform that state-space representation of the frequency-domain transfer matrix to one that is physically realisable. 
Applying the network synthesis theory can then lead to the physical setup of the quantum filter. 

This approach has powerful implications on how both passive and active quantum filters are designed, making the realisation of filters with arbitrarily complicated frequency responses a possibility. Since in principle we can view the entire measurement device as a many degrees-of-freedom quantum filter, this approach also provides a new paradigm for designing optimal quantum measurement devices. The outline of 
this paper goes as follows. In 
Section\,\ref{sec:approach}, we
present the mathematical details of this 
approach. In Section\,\ref{sec:unstable-filter}, 
we apply it to find a new optical 
realisation of the 
active unstable filter, the original proposal
of which was based upon optomechanics. 
We show that the optical loss is the limiting factor for the proposed optical 
realisation. In Section\,\ref{sec:discussion}, we summarise our result and provide 
an outlook of applying this approach to the
design of the optimal measurement devices. 

\section{Direct approach}
\label{sec:approach}

We now provide the details of the approach. The process to find a physical realisation, e.g.~an optical layout and its associated parameters, from a given set of transfer functions is general to multi-input multi-output lossless linear quantum systems; losses and other noise sources can be added later by augmenting the system description. Our starting point is the frequency-domain transfer function matrix, which is the square matrix that relates the frequency-domain system outputs $\mathbf{y}(s)$ to its inputs $\mathbf{u}(s)$:
\begin{align}
    G(s) &= C(-sI-A)^{-1}B + D,\label{eq:tf}\\ \mathbf{y}_i(s) &= \sum_j G_{ij}(s) \mathbf{u}_j(s),\nonumber
\end{align}
where $(A, B, C, D)$ are the system matrices as defined below, and $I$ is the identity matrix, and we assume that the number of inputs is equal to the number of outputs. Here the Laplace transform is defined as $f(s) = \int_{0^-}^{\infty} e^{+st} f(t) \mathrm{d}t$, with the lower bound at $t = 0^-$ so that an impulse can be added at $t = 0$. For a given transfer matrix a non-unique state-space representation can be found of the form~\cite{Luenberger1967, Ackermann1971, Kailath1980, Antoniou1988, Note5}:
\begin{align}
	\dot{\mathbf{x}} &= A\,\mathbf{x} + B\,\mathbf{u}, \label{eq:ss-1}\\
	\mathbf{y} &= C\,\mathbf{x} + D\,\mathbf{u}, \label{eq:ss-2}
\end{align}
which is a \emph{non-unique} time-domain representation of the system's dynamics. Here the quantity $\mathbf{x} \in \mathbb{L}^{2 n\times 1}$ ($\mathbb{L}$ being the space of linear operators on the relevant Hilbert space $\mathcal{H}$) is a vector of conjugate operator pairs representing the internal $n$ degrees of freedom of the system, $\mathbf{u} \in \mathbb{L}^{2m\times 1}$ is the vector of $m$ system inputs, and $\mathbf{y} \in \mathbb{L}^{2m\times 1}$ is the vector of $m$ system outputs. Note that in a quantum mechanical state-space, two conjugate operators are used to represent each individual degree of freedom of the system hence the factors of $2$. In the context of quantum optomechanics, $\hat{x}$ represents the cavity/oscillator eigenmodes for the cavities and mechanical oscillators in the system, while $\hat{u}$ and $\hat{y}$ are continuous Bosonic fields in free space~\cite{Blow1990,Nurdin2009}.
The dynamical matrix $A \in \mathbb{C}^{2n\times2n}$ describes the internal dynamics of the system, the input matrix $B \in \mathbb{C}^{2n\times2m}$ describes the coupling of the input into the system, the output matrix $C \in \mathbb{C}^{2m\times2n}$ describes the coupling of the system to the output, and the direct feedthrough matrix $D \in \mathbb{C}^{2m\times2m}$ describes the coupling of the input directly to the output. $(A, B, C, D)$ are together called the \emph{system matrices} and fully describe the linear dynamics of the system.

The system is called \emph{physically realisable} (and a corresponding physical realisation can be designed) if, in the Heisenberg picture evolution of the system, the commutation relations are preserved~\cite{James2008}:
\begin{equation}
	\forall i, j\ \ \ \mathrm{d}[\mathbf{x}_i,\mathbf{x}_j] = 0,\ \ [\mathbf{y}_i(t), \mathbf{y}^\dagger_j(t')] = \delta(t-t')\delta_{ij},
	\label{eq:physical-realisability-0}
\end{equation}
where the differential is treated using the quantum It\^o rule, meaning that the cross-products of the differentials of the operators must be calculated~\cite{Hudson1984,Parthasarathy1992,Bouten2006}.
The conditions on the system matrices for all such evolutions to preserve these commutation relations are found by using Eqs.\,\eqref{eq:ss-1} and\,\eqref{eq:ss-2} to calculate the increment of the system state $\mathrm{d}\mathbf{x}_i$ in Eq.\,\eqref{eq:physical-realisability-0} for an infinitesimal time period $\mathrm{d}t$. For an $n$ degree-of-freedom system described using complex mode operators (as are usually used in quantum optics) such that $\mathbf{x} = (\hat{a}_1, \hat{a}_1^\dagger;\dots;\hat{a}_n,\hat{a}_n^\dagger)^T$ with $m$ inputs and outputs described by $\mathbf{u} = (\hat{u}_1, \hat{u}_1^\dagger;\dots;\hat{u}_m,\hat{u}_m^\dagger)^T$ and similarly for $\mathbf{y}$, the constraints on the system matrices are given by
\begin{align}
	A J + J A^\dagger + B J_m B^\dagger &= 0, 
	\label{eq:physical-realizability-1}\\
	J C^\dagger + B J_m D^\dagger &= 0, \label{eq:physical-realizability-2}\\
	\ \ D J_m D^\dagger &= J_m,
	\label{eq:physical-realizability-3}
\end{align}
where $J = \text{diag}(1, -1; \dots; 1, -1)\in \mathbb{R}^{2n \times 2n}$ and $J_m = \text{diag}(1, -1; \dots; 1, -1)\in \mathbb{R}^{2m \times 2m}$\footnote{As discussed in Section.~\ref{appendix:r-matrix} of this paper, this matrix takes a different form when using Hermitian observable quadrature operators.}. See Appendix A of~\cite{Gough2009} for a proof of these constraints. So now we have a restriction on the possible system matrices that can lead to a physically realisable system.

Now we consider how to generate such a physically-realisable state-space model from the system's transfer matrix. The conventional procedure for transforming the transfer matrix to a minimal state-space model is outlined in Refs.\,~\cite{Luenberger1967, Kailath1980}. Such a state-space model constructed from a pole-zero form transfer matrix is minimal if the number of internal degrees of freedom (e.g.~the number of pairs of conjugate ladder operators describing the system state) is equal to the highest polynomial order in the frequency $s$ among all of the transfer functions in the transfer matrix. Generally such a procedure will lead to system matrices $(A',B',C',D')$ that do not satisfy Eqs.~\eqref{eq:physical-realizability-1} and~\eqref{eq:physical-realizability-2} and therefore cannot be physically realised. 

Here we show a method allowing us to transform a minimal realisation $(A',B',C',D')$ to a physically realisable counterpart $(A, B, C, D)$, given that the transfer matrix ${G}(s)$ obeys a condition that will be given in Eq.~\eqref{eq:j-j-unitary} and the state-space realisation satisfies the conditions given in \cite[Theorem 3]{Shaiju2012} (to be recalled below). The transformation is achieved by looking for a Hermitian matrix $X \in \mathbb{C}^{2n\times2n}$ that obeys the constraints:
\begin{align}
	A' X + X (A')^\dagger + B' J_m (B')^\dagger &= 0,
	\label{eq:physical-realizability-X1} \\
	X (C')^\dagger + B' J_m (D')^\dagger &= 0.
	\label{eq:physical-realizability-X}
\end{align}
This matrix $X$ can be written in the form of a similarity transformation $X = TJT^\dagger$ for some non-singular matrix $T$. Substituting this into Eqs.~\eqref{eq:physical-realizability-1} and~\eqref{eq:physical-realizability-2} we see that the conditions are satisfied after making the transformations,
\begin{equation}
A = T^{-1} A' T,\ B = T^{-1} B',\ C = C' T,\ D = D',
\label{eq:applying-similarity-transformation}
\end{equation}
to find the physically realisable state-space $(A, B, C, D)$. The existence of $X$, i.e. $T$, is guaranteed by the 
symplectic condition imposed on any physically realisable transfer matrix ${G}(s)$ and direct-feed matrix $D$\,\cite{Shaiju2012} (where $\null^*$ denotes the complex conjugate of a complex number):
\begin{equation}
    {G}^\dag(s^*)J_m{G}(-s) = J_m, 
    \label{eq:j-j-unitary}
\end{equation}
and fulfilment of the conditions of \cite[Theorem 3]{Shaiju2012}, which says that $\lambda_i(A)+\lambda_j(A)^* \neq 0$ for any pair of eigenvalues $\lambda_i(A)$ and $\lambda_j(A)$ of $A$, and the feedthrough matrix $D$ is unitary and satisfies \eqref{eq:physical-realizability-3}.

Now that we have shown how to obtain the physically realisable state-space model $(A, B, C, D)$ from the transfer matrix obeying Eq.~\eqref{eq:j-j-unitary} we can infer the physical realisation. We describe the realisation chiefly using the generalised open oscillator~\cite{Nurdin2009} formalism. This is a general formalism describing open quantum systems with arbitrary internal linear dynamics and input-output couplings, providing a language for describing and analysing systems with internal degrees of freedom coupled to external continuum fields, such as quantum measurement devices and quantum filters. As shown in~\cite{James2008} for an $n$ degree-of-freedom system, when the direct-feed matrix $D$ is symplectic and unitary (i.e.~it satisfies \eqref{eq:physical-realizability-3} and $D^{\dag}D = DD^{\dag}=I$), there is a one-to-one correspondence between the system matrices $(A, B, C, D)$ and the generalised open oscillator which is parameterized by a triplet $(S, \hat{L}, \hat{H})$~\cite{Nurdin2009, Tezak2012a, Combes2016}. Here, the scattering matrix $S \in \mathbb{C}^{m \times m}$ describes the transformation of the input fields through a passive network, i.e.~any passive pre-processing of the system's input fields. The coupling operator $\hat{L} = K \mathbf{x}$ where $K \in \mathbb{C}^{m\times 2n}$ describes the coupling between the input and output fields and the internal degrees of freedom, e.g.~equivalent to the usual input-output Langevin equations~\cite{Chen2013} when the input-output fields are coupled to the internal fields by a mirror. The Hamiltonian $\hat{H}$ describes the free evolution of the internal system dynamics as if the system were closed. The relation between the system matrices and the generalised open oscillator parameters is given by,
\begin{align}
\label{eq:SKR}
\begin{split}
	S_{kl} &= D_{2k-1,2l-1}\,, \quad 
	\hat{L} = [\begin{array}{cc} I & 0 \end{array}]P\, C\, \mathbf{x},
	\\ \hat{H} &= \frac{i}{4} \hbar\,\mathbf{x}^\dag  \left(J A - A^\dagger J\right) \mathbf{x},
\end{split}
\end{align}
where $I$ is the identity matrix and $0$ is the null matrix, both of dimension $m$, $P$ is the permutation matrix that maps $\mathbf{u} = (\hat{u}_1,\hat{u}_1^{\dag};\ldots;\hat{u}_m,\hat{u}_m^{\dag})^T$ to $(\hat{u}_1,\hat{u}_2,\ldots,\hat{u}_m,\hat{u}_1^{\dag},\hat{u}_2^{\dag},\ldots,\hat{u}_m^{\dag})^T$, and $\hat{H}$ is derived in Section.~\ref{appendix:r-matrix}. The total Hamiltonian is then
\begin{equation}
    \hat{H}_\text{tot} = \hat{H} + i\hbar[\begin{array}{cc} \hat{L}^{\dag} & -\hat{L}^{T} \end{array}] \mathbf{u},
\end{equation}
where the input fields $\mathbf{u}$ are pre-processed by a static passive network described by $S$. Now we have achieved the full physical Hamiltonian describing the system starting from the transfer matrix describing the frequency-domain input-output behaviour.

In the case where only one of the internal degrees of freedom is coupled to the input-output fields, it could be straightforward to construct the physical realisation by inspection, as in the  illustrative example of the unstable filter
discussed in the next section. However, having one internal degree of freedom is not always the case. Systems consisting of more than one internal degree-of-freedom can first be sub-divided into separate one degree-of-freedom systems coupled via direct interaction Hamiltonians via the main synthesis theorem proved in~\cite{Nurdin2009}. These systems can then be systematically realised by connecting the individual one degree-of-freedom systems in series, and overlapping them accordingly, giving a systematic way to construct the physical realisation regardless of complexity~\footnote{Note that the approach is also entirely general to optomechanical systems, provided that the dynamics can be linearised.}. 
The outline of such a general approach to constructing the physical realisation given an $n$ degree-of-freedom generalised open oscillator goes as follows:
\begin{enumerate}
    \item First, the main synthesis theorem is used to split the $n$ degree-of-freedom oscillator into $n$ one degree-of-freedom oscillators which are connected in series, i.e.~the output of each oscillator is fed into the input of the next for example via a beamsplitter which is known as the series product~\cite{Gough2007}, and also a direct interaction Hamiltonian is produced coupling the oscillators. Often the series product connection is not needed as only one of the internal modes is coupled to the external continuum, i.e.~the operator $\hat{L}$ is only non-zero for one of the internal modes. The task is then to realise each of these one degree-of-freedom oscillators and the direct interaction Hamiltonian.
    \item For each one degree-of-freedom oscillator we do the following. First the scattering matrix can be realised as a static passive linear network using only beamsplitters and mirrors. Then, the general coupling operator of the form $\hat{L} = \alpha \hat{a} + \beta \hat{a}^\dag$ can be realised by indirectly coupling the mode $\hat{a}$ to the external continuum fields $\hat{u}$ and $\hat{y}$ via an auxiliary mode $\hat{b}$, which has sufficiently fast dynamics with coupling rate $\gamma$ such that it can be adiabatically eliminated from the final input-output relation. This auxiliary mode is coupled to the main mode via non-linear crystal (two-mode squeezing process) for the $\beta \hat{a}^\dagger$ term, and via a beamsplitter for the $\alpha \hat{a}$ term. These are related to the physical parameters via $\alpha = - \epsilon_2^* \sqrt{2/\gamma}$ where $\epsilon_2 = 2 \Theta e^{-i\Phi}$ where $\Theta$ is the beamsplitter mixing angle and $\Phi$ is the relative phase detuning introduced by the beamsplitter, and $\beta = \epsilon_1 \sqrt{2/\gamma}$ where $\epsilon_1$ is the effective pump intensity, shown to be equal to $c r / (2L)$ in Section.~\ref{appendix:single-pass}, where $r$ is the single-pass squeezing factor and $L$ is the cavity length. The resulting interaction Hamiltonian is $\hat{H}_{ab} = \hbar (\epsilon_1 \hat{a}^\dagger \hat{b}^\dagger + \epsilon_1^* \hat{a} \hat{b}) + \hbar (\epsilon_2 \hat{a}^\dagger \hat{b} + \epsilon_2^* \hat{a} \hat{b}^\dagger)$. Finally, the Hamiltonian $\hat{H}$ can be realised in the most general case as a detuned DPA (degenerate parametric amplifier), which can be implemented as a detuned cavity with a $\chi^{(2)}$ non-linear crystal with a pump frequency twice the laser carrier frequency $\omega_0$. Specifically to realise the Hamiltonian $\hat{H} = \hbar \Delta \hat{a}^\dagger \hat{a}^{\vphantom{\dagger}} + \hbar (\epsilon (\hat{a}^\dag)^2 + \epsilon^* \hat{a}^2)$ we use a cavity with resonant frequency $\omega_\text{cav} = \omega_0 + \Delta$ where $\omega_0$ is the laser carrier frequency and a non-linear crystal with effective pump intensity $\epsilon = c r / (2L)$ where again $r$ is the single-pass squeezing factor and $L$ is the cavity length.
    \item To implement the interaction Hamiltonian between each one degree-of-freedom oscillator we overlap the relevant internal modes of each oscillator via a non-linear crystal and/or beamsplitter depending on the interaction. The interaction Hamiltonian between modes $\hat{a}_k$ and $\hat{a}_l$ can be written in the form $\hat{H}_{kl} = \hbar(\epsilon_2 \hat{a}^\dag_k \hat{a}_l + \epsilon_2^* \hat{a}_k \hat{a}^\dag_l + \epsilon_1 \hat{a}^\dag_k \hat{a}^\dag_l + \epsilon_1 \hat{a}_k \hat{a}_l)$ where in this case the effective pump intensity is $-2i\epsilon_1$, and again $\epsilon_2 = 2 \Theta e^{-i\Phi}$ where $\Theta$ is the mixing angle and $\Phi$ is the relative phase difference between the two modes. In the unstable filter example below we do not need to do this as we already only have one internal degree of freedom.
\end{enumerate}

\begin{figure}[!b]
    \centering
    \includegraphics[width=0.95\linewidth]{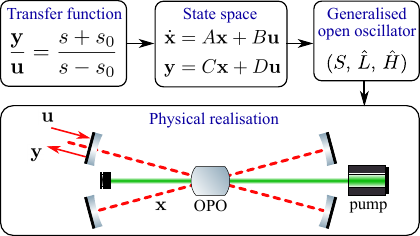}
    \caption{Flowchart showing the steps in constructing the physical realisation of a quantum filter; an active filter is used as an illustration. }
    \label{fig:realisation_steps}
\end{figure}

In summary, from the input-output transfer matrix we have developed the physical parameters describing the system. The procedure is summarised in Fig.\,\ref{fig:realisation_steps}.

\section{Illustrative example: an unstable filter}
\label{sec:unstable-filter}

To demonstrate the power of this approach, we will go beyond passive optical cavities by considering a non-trivial active filter for beating the universal gain-bandwidth product limit in resonant detection schemes~\cite{Wicht1997, Muller2000, Wise2004, Pati2007, Yum2013, Ma2015, Miao2015, David2015, Miao, Page2018, Bentley2019, Zhou2015, Zhou2018, Page2019, Shimazu2019}, specifically, the so-called unstable filter~\cite{Miao2015} which has a broadband anomalous (negative) dispersion. The gain-bandwidth product, also known as the Mizuno limit~\cite{JunMizuno}, states that the integral of the squared frequency-domain signal transfer function for a resonant detector is bounded purely by the energy stored in the detector. Therefore, when considering the quantum shot noise due to the input quantum vacuum, the detection bandwidth and peak sensitivity are inversely proportional which cannot be surpassed by changing any physical parameter other than the power in the detector. One way this can be surpassed is by using the broadband anomalous dispersion of the aforementioned unstable filter to partially negate the positive dispersion of the detector's signal cavity as first discussed in~\cite{Miao2015}. The filter also is unusual because it seemingly violates the Kramers-Kronig relations which imply that a stable anomalous dispersion filter without absorption (i.e.~with unity gain over the range of the anomalous dispersion) violates causality, however since this system is dynamically unstable this restriction does not apply~\cite{Kronig1926, Toll1956a, Doyle1990, Hirschorn2009}. As discussed in the procedure above we start with the frequency-domain transfer function of the unstable filter,
\begin{equation}
    G(s) = \frac{s - s_0}{s + s_0},
    \label{eq:transfer-function}
\end{equation}
where $s \equiv i \omega $ and $s_0 = \gamma_\text{neg}$ is a characteristic frequency quantifying the anomalous (negative) dispersion. An an example will now infer a physical realisation for this device using the above procedure.

First we note that since the transfer function is first order in frequency $s$, only one internal degree of freedom will be required for the minimal state-space realisation. Therefore the system state vector $\hat{x}$ will have two elements: $\mathbf{x}_1 = \hat{a}, \mathbf{x}_2 = \hat{a}^\dagger$ describing a single cavity mode, and similarly for the vectors $\mathbf{u}$ and $\mathbf{y}$ describing the input and output modes respectively. As a transfer-function 
matrix, Eq.~\eqref{eq:transfer-function} can be written as
\begin{equation}
    \mathbf{G}(s) = \frac{s-s_0}{s+s_0} \begin{bmatrix} 1 & 0 \\ 0 & 1\end{bmatrix},
\end{equation}
which can be verified to satisfy the constraint \eqref{eq:j-j-unitary} and therefore a corresponding physical realisation can be found. To simplify the notation, we define a dimensionless $s$ (and the corresponding time) which is normalised with respect to $s_0/2=\gamma_{\rm neg}/2$ (a factor of $2$ for convenience), namely $s\rightarrow (s_0/2)s$ where $s$ is now dimensionless.
A corresponding minimal but not physically realisable state-space model is given by,
\begin{align}
	\begin{bmatrix} \dot{\hat{a}} \\ \dot{\hat{a}}^\dagger \end{bmatrix}
	&=
	\begin{bmatrix} 2 & 0 \\ 0 & 2 \end{bmatrix}
	\begin{bmatrix} \hat{a} \\ \hat{a}^\dagger \end{bmatrix}
	+
	\begin{bmatrix} \hat{u} \\ \hat{u}^\dagger \end{bmatrix},\\
	\begin{bmatrix} \hat{y} \\ \hat{y}^\dagger \end{bmatrix}
	&=
	\begin{bmatrix} 4 & 0 \\ 0 & 4 \end{bmatrix} 
	\begin{bmatrix} \hat{a} \\ \hat{a}^\dagger \end{bmatrix}
	+
	\begin{bmatrix} \hat{u} \\ \hat{u}^\dagger \end{bmatrix}. 
\end{align}

The matrix $X$ that solves Eqs.~\eqref{eq:physical-realizability-X1} and~\eqref{eq:physical-realizability-X} is given by $X = -J / 4$, which can be written in the form $X = TJT^\dag$ with the matrix $T$ which transforms the above state-space model to the physically realisable one is given by,
\begin{equation}
	T = \frac{1}{2} \begin{bmatrix} 0 & -1 \\ 1 & 0 \end{bmatrix}.
\end{equation}
The resulting dimensionless state-space model can be found by applying the similarity transformation as shown in Eq.~\eqref{eq:applying-similarity-transformation},
\begin{equation}
	\tilde{A} = \begin{bmatrix}2 & 0 \\ 0 & 2\end{bmatrix},\ 
	\tilde{B} = \begin{bmatrix}0 & 2 \\ -2 & 0\end{bmatrix},\ 
	\tilde{C} = \begin{bmatrix}0 & -2 \\ 2 & 0\end{bmatrix},\ \tilde{D} = I,
	\label{eq:state-space}
\end{equation}
which obey Eqs.\,\eqref{eq:physical-realizability-1} and \eqref{eq:physical-realizability-2} by construction and therefore is a physically realisable state-space model. Reversing the normalisation process gives us,
\begin{align}
	A &= \begin{bmatrix}s_0 & 0 \\ 0 & s_0\end{bmatrix},\ 
	B = \begin{bmatrix}0 & \sqrt{2 s_0} \\ -\sqrt{2 s_0} & 0\end{bmatrix},\nonumber\\
	C &= \begin{bmatrix}0 & -\sqrt{2 s_0} \\ \sqrt{2 s_0} & 0\end{bmatrix},\ D = \tilde{D},
	\label{eq:state-space-with-dimensions}
\end{align}

Eq.\,\eqref{eq:SKR} can now be used to calculate the scattering matrix, input-output coupling, and internal Hamiltonian for the unstable filter. We have
\begin{equation}
    S = I,\ \hat{L} = -\sqrt{2 s_0}\hat{a}^\dag,\ \hat{H}=0.
\end{equation}
This implies that there is no input scattering with $S = I$, and $\hat{L} = -2\hat{a}^\dagger$, and there is no detuning or internal squeezing of the cavity mode as $\hat{H}=0$.

\begin{figure}[t]
    \centering
    \includegraphics[width=\linewidth]{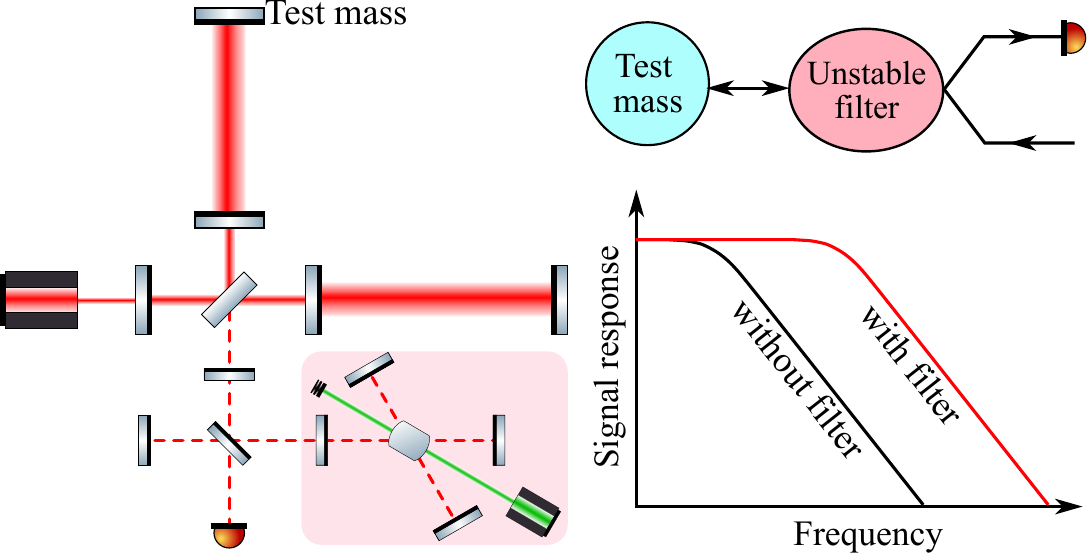}
    \caption{Diagram showing where the filter realisation (highlighted by the shaded box) would be integrated into a standard Michelson inteferometer, using a scheme similar to that proposed in Ref.\,\cite{Miao2015}, which improves the signal response at high frequencies via the negative dispersion compensating the positive dispersion of the arm cavities as discussed above.}
    \label{fig:integrated-in-setup}
\end{figure}

Since $\hat{H}$ and $S$ are trivial to implement we now implement the coupling operator $\hat{L} = -\sqrt{2 s_0}\hat{a}^\dag$ as discussed above. In this case we have $\alpha = 0$, and so just have the auxiliary mode coupled to the main cavity mode via a non-linear crystal. This auxiliary mode will later be adiabatically eliminated, however it makes the physical realisation more feasible as coupling two cavity modes via a parametric oscillator is more experimentally durable. Therefore we construct the physical realisation shown in Fig.\,\ref{fig:realisation_steps}, which can be integrated into an interferometer as shown in Fig.\,\ref{fig:integrated-in-setup}. 

The realisation simply consists of two tuned cavities (the main mode $\hat{a}$ and the auxiliary mode $\hat{b}$) coupled via a $\chi^{(2)}$ non-linear crystal, labelled OPO (optical parametric oscillator), pumped by a classical pump field, labelled pump. One of the cavities is coupled to the external fields. Specifically, we have
\begin{align}
	\hat{H}_{ab} & = -\hbar \sqrt{s_0\,\gamma} (\hat{a}^\dagger \hat{b}^\dagger + \hat{a}\hat{b})\,, 
	\label{eq:interaction-hamiltonian-2}\\
	\hat H_{\rm ext} & = - i \hbar \sqrt{\gamma} 
	(\hat b\,\hat c_{\rm ext}^{\dag} - \hat b^{\dag} \hat c_{\rm ext})\,.
	\label{eq:interaction-hamiltonian-external-continuum}
\end{align}
The interaction Hamiltonian $\hat H_{ab}$ describes the coupling of both cavity modes $\hat{a}$ and $\hat{b}$ via the OPO. As shown in Section.~\ref{appendix:single-pass}, the coupling rate $\sqrt{s_0\,\gamma}$ is equal to $rc/(2L_b)$, where $r$ is the  single-pass squeezing factor of the crystal and $L_b$ is the length of the auxiliary  cavity. As an order of magnitude estimate for implementation in a laser interferometer with arm length of $L_\text{arm} = \SI{4}{\kilo\metre}$ (where $s_0 \equiv \gamma_\text{neg} = c/L_\text{arm}$~\cite{Miao2015}), the required squeezing factor is
\begin{equation}
    r = 7.7\times 10^{-5} \sqrt{\frac{T_b}{\SI{100}{\ppm}}} \sqrt{\frac{L_b}{\SI{24}{\centi\metre}}\vphantom{\frac{T_b}{\SI{10}{\ppm}}}} \sqrt{\frac{\SI{4}{\kilo\metre}}{L_\text{arm}}\vphantom{\frac{T_b}{\SI{10}{\ppm}}}}.
\end{equation}
The Hamiltonian $\hat H_{\rm ext}$ describes the coupling
between the auxiliary mode $\hat b$ and the external
continuum field $\hat c_{\rm ext}$, which is related 
to the input and output operators via $\hat u\equiv \hat c_{\rm ext}(t=0_-)$ and $\hat y\equiv \hat c_{\rm ext}(t=0_+)$~\cite{Walls2008, Chen2013}. The coupling rate $\gamma$ is defined as $T_b c / (4 L_b)$
where $T_b$ is the input mirror transmissivity.
The negative dispersion transfer function shown in Eq.~\eqref{eq:transfer-function} can then be recovered 
by solving the resulting Heisenberg equations of motion 
in the frequency domain, and then applying the approximation $\gamma \gg \omega$, the so-called ``resolved-sideband regime'', which effectively adiabatically eliminates $\hat{b}$~\cite{Nurdin2009}.

In Section.~\eqref{appendix:losses}, we include the 
effect of optical loss for the realistic implementation. 
We found that the noise contribution from the auxiliary cavity loss is insignificant compared to the contribution from the $\hat{a}$ cavity loss. The resulting 
input-output relation including the optical loss is given by
\begin{equation}
    \hat{y}(s) \approx \frac{\omega + i(\gamma_a^\epsilon + s_0)}{\omega + i(\gamma_a^\epsilon - s_0)} \hat{u}(s) + \frac{2 \sqrt{s_0\,\gamma_a^\epsilon}}{\omega + i(\gamma_a^\epsilon - s_0)} \hat{n}_a^\dagger(s),
\end{equation}
where $\gamma_a^\epsilon = \epsilon_a c / (4 L_a)$ with $\epsilon_a$ being the total optical loss in the $\hat{a}$ cavity and $L_a$ being the cavity length, and $\hat{n}_a$ is 
the corresponding vacuum noise process. The distortion of the transfer function due to $\gamma_a^\epsilon$ is on the order of $\gamma_a^\epsilon / s_0$, while the noise term is on the order of $\sqrt{\gamma_a^\epsilon / s_0}$ and is therefore more significant. 

\begin{figure}[t]
    \centering
    \includegraphics[width=\linewidth]{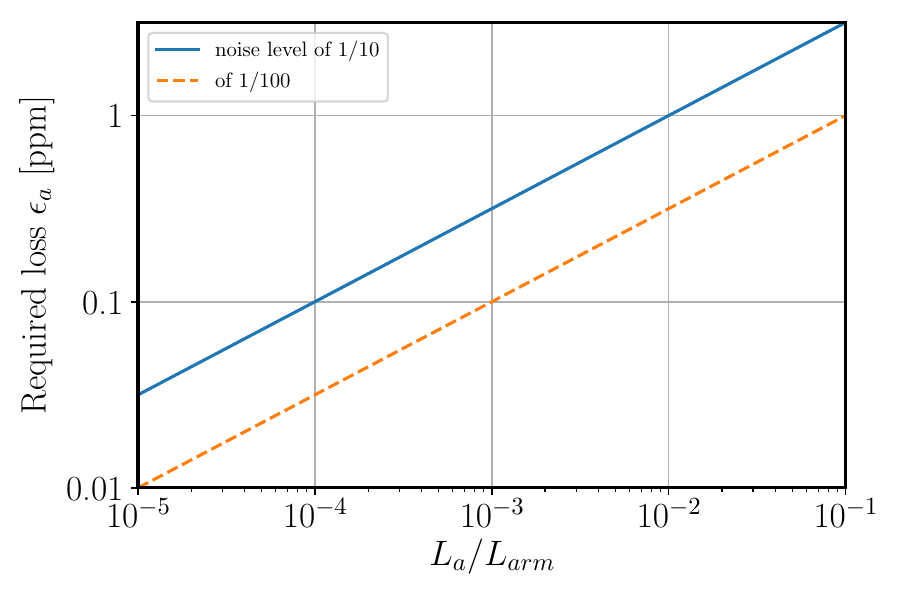}
    \caption{Required total $\hat{a}$ cavity loss $\epsilon_a$ as a function of ratio of $\hat{a}$ cavity length to arm cavity length $L_a / L_\text{arm}$ for the cases where the noise power contribution at $\omega = 0$ due to $\hat{n}_a$ is a tenth of that of the signal power (blue line) and a hundred (orange line).}
    \label{fig:required_losses}
\end{figure}

The above input-output relation takes the same form as the optomechanical case~\cite{Miao2015} if we view $\hat n_a$ 
as the thermal noise of the mechanical oscillator. 
In contrast, in this case the loss $n_a$ is sourced by the quantum vacuum and so it only has vacuum fluctuations, equivalent to a mechanical oscillator at environmental temperature $T_\text{env} = 0$. Therefore the strict thermal requirements of the optomechanical unstable filter are avoided. Instead vacuum fluctuations are injected due to losses in the mirrors and the non-linear crystal. The required loss to achieve low noise as a function of $\hat{a}$ cavity length is shown in Fig.~\ref{fig:required_losses}. As we can see, given an interferometer arm length of $L_\text{arm} = \SI{4}{\kilo\metre}$, a loss per unit length of $\epsilon_a / L_a = \SI{2.5}{\ppm\per\metre}$ is required to achieve a $1/10$ noise contribution, which is already achievable with state-of-the-art optics for the free space part of the cavity~\cite{Isogai2013,Oelker2016a}. The loss of a \SI{1}{\centi\metre} crystal is between around 15--40 \si{\ppm} \cite{Bogan2015}, so taking a middle value of around $\SI{25}{\ppm}$ then to achieve a loss per unit length of $\SI{2.5}{\ppm\per\metre}$ the cavity length must be at least \SI{10}{\metre}. Further assuming losses of \SI{10}{\ppm} for each side of the anti-reflective coating of the crystal drives the minimum cavity length up to \SI{18}{\metre}.

\section{Discussion}
\label{sec:discussion}

In general the physical realisation produced may have unstable internal dynamics, as is the case for the unstable filter above. Since the single-mode approximation is used, the stabilising controller previously derived in~\cite{Miao2015} can be used for each unstable degree of freedom.

In addition to 
realising quantum filters with a known transfer function, this approach can also be used to design the optimal high-precision measurement devices, where the optimality is based upon the quantum Cram\'{e}r-Rao bound\,~\cite{Helstrom1967,Holevo2011,Braunstein1996,Giovannetti2011,Tsang2011,Miaoa}. We can view the
entire measurement device as a $N$ degree-of-freedom quantum filter, and then tune the filter parameters so as to minimise the quantum Cram\'{e}r-Rao bound. Therefore, we can construct the most sensitive possible $n$ degree-of-freedom
measurement device. This opens up a new paradigm of designing and optimising measurement devices and is worthy of being further explored.

\section{Acknowledgements}

We would like to thank Rana Adhikari, Denis Martynov, Naoki Yamamoto, LIGO AIC, and QNWG for fruitful discussions. 
J.B. is supported by STFC and School of Physics and
Astronomy at the University of Birmingham. J.B. and H.M.
acknowledge the additional support from the Birmingham
Institute for Gravitational Wave Astronomy.
H.M. has also been supported by UK STFC Ernest Rutherford 
Fellowship (Grant No. ST/M005844/11). Y.C. is supported by the Simons Foundation (Award Number 568762), and the National Science Foundation, through Grants PHY-1708212 and PHY-1708213. 

\appendix

\section{\label{appendix:r-matrix}Hamiltonian matrix in complex operator notation}

In this section the expression for the internal Hamiltonian $\hat{H}$ in terms of the system matrices will be transformed from the real-quadrature form in Ref.~\cite{James2008} to the complex ladder operator form in Eq.~\eqref{eq:SKR}.

The Hamiltonian in the real-quadrature form is given by 
\begin{equation}
    \hat{H} = \mathbf{x}_r^\dagger \Omega_r \mathbf{x}_r\,, 
\end{equation}
where $\mathbf{x}_r = (\hat{q}_1, \hat{p}_1;\dots;\hat{q}_n, \hat{p}_n)^T$ are the real quadrature operators. 
The relation between $\Omega_r$ and 
the dynamical matrix $A_r$ in the state-space model
is given uniquely by,
\begin{equation}
    \Omega_r = \frac{1}{4} \left(-\Theta A_r + A_r^\dagger \Theta\right),
\end{equation}
where,
\begin{equation}
    \Theta = \text{diag}(\underbrace{\Theta_1,\dots,\Theta_1}_{\text{n times}}) \in \mathbb{R}^{2n\times2n},
\end{equation}
and,
\begin{equation}
    \Theta_1 = \begin{bmatrix}0 & 1 \\ -1 & 0\end{bmatrix}.
\end{equation}

The complex ladder operators are related to the real quadrature operators by $\mathbf{x} = (\hat{a}^{\vphantom{\dagger}}_1, \hat{a}_1^\dagger;\dots;\hat{a}^{\vphantom{\dagger}}_n, \hat{a}_n^\dagger)^T = U \mathbf{x}_r$, where,
\begin{equation}
	U = \text{diag}(\underbrace{U_1,\dots,U_1}_{\text{n times}}) \in \mathbb{C}^{2n\times2n},
\end{equation}
where,
\begin{equation}
    U_1 = \frac{1}{\sqrt{2}} \begin{bmatrix}
		1 & i \\ 1 & -i
	\end{bmatrix},
\end{equation}
is the unitary transformation that converts from the real quadrature operators $(\hat{q}, \hat{p})$ to the complex ladder operators $(\hat{a}, \hat{a}^\dagger)$.

Note that we can write $\Theta = -iU^\dagger J U$, and that the relation between the dynamical matrix in the real quadrature picture and the complex ladder operators is given by $A = U^\dagger A_r U$, and recall that $U$ is unitary. Substituting these facts into the expression for $\hat{H}$ we get $\hat{H} = \mathbf{x}^\dagger \Omega \mathbf{x}$ where,
\begin{equation}
    \Omega = \frac{i}{4}\left(JA-A^\dagger J\right),
\end{equation}
where $J = \text{diag}(1,-1;\dots;1,-1) \in \mathbb{R}^{2n\times2n}$.

\section{\label{appendix:single-pass}Relating the coupling rate to the single-pass squeezing factor}

To compare the coupling rate $\sqrt{s_0 \gamma}$ to the single-pass amplification factor $r$, we look at the degenerate case of the interaction Hamiltonian given Eq.~\eqref{eq:interaction-hamiltonian-2},
\begin{equation}
     \hat H_{\rm deg}= -\hbar \sqrt{s_0 \gamma}[(\hat{a}^\dagger)^2 + \hat{a}^2].
\end{equation}
Solving the equation of motion in the 
frequency domain, the resulting input-output relation 
for the amplitude quadrature $\hat a_1$ in the 
two-photon formalism~\cite{Caves1985,Schumaker1985} is 
\begin{equation}\label{eq:app_io1}
    \hat a_1^{\rm out}(\omega) = \frac{\gamma+\sqrt{s_0\gamma} + i\omega} {\gamma-\sqrt{s_0\gamma} - i\omega}\hat a_1^{\rm in}(\omega)\,. 
\end{equation}

We can derive the same input-output relation by 
propagating the continuum field through the cavity with 
a nonlinear crystal, and 
obtain 
\begin{equation}
     \hat a_1^{\rm out}(\omega)  = \frac{-\sqrt{R} + e^{2r}e^{2i \omega L/c}}{1-\sqrt{R}\,e^{2r}e^{2i \omega L/c}} \hat a_1^{\rm in}(\omega)\,. 
\end{equation}
Assuming $T\equiv 1-R, r, \omega L/c \ll 1$, we can make the 
Taylor expansion of the above equation to the leading 
order of these small dimensionless quantities:  
\begin{equation}\label{eq:app_io2}
    \hat a_1^{\rm out}(\omega)  \approx \frac{T / 2 + 2r + 2i\omega L/c}{T / 2- 2r  -2i\omega L/c}\hat a_1^{\rm in}(\omega)\,.
\end{equation}
Eq.\,\eqref{eq:app_io1} and Eq.\,\eqref{eq:app_io2} become
identical when 
\begin{equation}
    \gamma\equiv \frac{Tc}{4 L}\,,\quad r = 2 \sqrt{s_0\gamma }\,\frac{ L}{c}\,,  
\end{equation}
which is the mapping used following Eq.~\eqref{eq:interaction-hamiltonian-external-continuum}. 

\section{\label{appendix:losses}Including losses into the analysis}

In this section, we show how the effect of optical loss 
is included in the analysis for the realistic implementation. 
The optical losses in the mirrors of both cavities will introduce quantum white noise vacuum processes~\cite{Gardiner1991,Braginsky,Nurdin2009}, $\hat{n}_a, \hat{n}_b$, which are coupled to modes $\hat{a}$ and $\hat{b}$ respectively via transmissivities $T_a, T_b$. This results in extra terms added to the Heisenberg equations of motion for the two modes,
\begin{align}
	\dot{\hat{b}} &= -\gamma_b^\epsilon \hat{b} + \sqrt{2 \gamma_b^\epsilon} \hat{n}_b + \frac{i}{\hbar} [\hat{H}_\text{tot},\hat{b}] , \label{eq:losses-b} \\
	\dot{\hat{a}} &= -\gamma_a^\epsilon \hat{a} + \sqrt{2 \gamma_a^\epsilon} \hat{n}_a + \frac{i}{\hbar} [\hat{H}_\text{tot},\hat{a}] \label{eq:losses-a},
\end{align}
where $H_\text{tot}$ is the total Hamiltonian given in Eqs.~\eqref{eq:interaction-hamiltonian-2} and~\eqref{eq:interaction-hamiltonian-external-continuum}. The noise coupling constants for the $\hat{a}$ cavity and $\hat{b}$ cavity respectively are given by:
\begin{equation}
\gamma_a^\epsilon = \epsilon_a c / (4 L_a)\,,\quad \gamma_b^\epsilon = \epsilon_b c / (4 L_b),
\end{equation}
where $\epsilon_a$ and $\epsilon_b$ are the optical losses described by cavity respectively. The loss from the non-linear crystal couples identically to the mirror loss into both cavities, and so can be included in $\epsilon_a, \epsilon_b$.

Solving the Heisenberg equations of motion in the frequency domain, we found that the noise contribution from the auxiliary cavity loss $\hat{n}_b$ is much smaller than the contribution from the $\hat{a}$ cavity loss $\hat{n}_a$ by a factor:
\begin{equation}
\frac{\omega^2 \gamma_b^\epsilon} {\gamma_\text{neg} \gamma \gamma_a^\epsilon} \ll 1,
\end{equation}
assuming $\gamma_a^\epsilon \approx \gamma_b^\epsilon$, and $\omega \ll \gamma_\text{neg}$, $\omega \ll \gamma$, a result also found in the optomechanical case explored in~\cite{Miao2015}, in which the filter cavity takes the role of the auxiliary cavity mode $\hat{b}$ and the mechanical oscillator takes the role of the main cavity mode $\hat{a}$. However in our case the main cavity loss is due to vacuum and is not thermally driven, and so is effectively at zero temperature. The phase noise due to the thermal fluctuation of the non-linear crystal~\cite{Cesar2009} is negligible as there is almost no carrier power in either cavity.

\section{Alternative topology}

\begin{figure}[ht]
    \centering
    \includegraphics[width=0.9\linewidth]{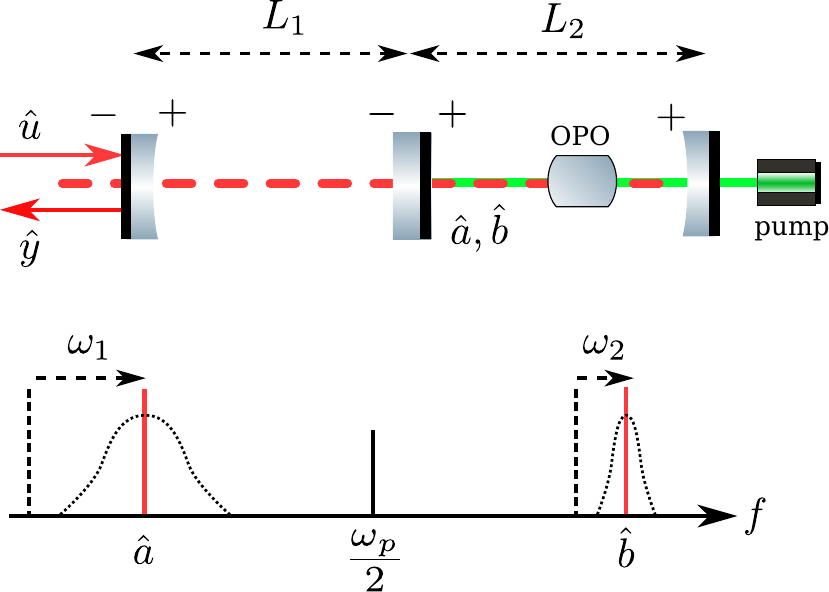}
    \caption{Optical diagram and relevant frequencies of the alternative topology, consisting of a non-linear crystal and two linear cavities with the crystal in only one cavity.}
    \label{fig:twice-FSR-realization}
\end{figure}

In Fig.~\ref{fig:twice-FSR-realization} we show an alternative topology for the realisation shown in Fig.~\ref{fig:realisation_steps}. The system consists of a linear coupled cavity. We call the cavity with 
the nonlinear crystal in it the active cavity and the other the passive cavity. The length of the passive cavity $L_1$ differs from the length $L_2$ of the active cavity so that they have different mode spacings. The two modes $\hat a$ and $\hat b$ in this case belong to the same longitudinal modes of the active cavity but separated by one free spectral range. The passive cavity acts as a compound mirror with frequency-dependent effective phase $\phi_\text{eff}(\omega)$ and transmissivity $T_\text{eff}(\omega)$, the former shifting the resonances of the active cavity by $\omega_a$ and $\omega_b$ for the $\hat{a}$ and $\hat{b}$, and the latter imparting different bandwidths for the two modes, denoted $\gamma_a = T_\text{eff}(\omega_a) c / (4 L_2)$ and $\gamma_b = T_\text{eff}(\omega_b) c / (4 L_2)$ respectively. The non-linear crystal pump frequency is set to $\omega_p$ where $\omega_p / 2$ is between the two modes $\hat{a}$ and $\hat{b}$. To make $\hat b$ satisfy the adiabatic condition, we require $\gamma_b \gg \omega$, while to ensure good performance we require $\gamma_a \ll \gamma_\text{neg}$. Both bandwidths can be independently controlled by changing the relative lengths of the two cavities.

\bibliography{bibliography-processed.bib}

\begin{thebibliography}{76}%
\makeatletter
\providecommand \@ifxundefined [1]{%
 \@ifx{#1\undefined}
}%
\providecommand \@ifnum [1]{%
 \ifnum #1\expandafter \@firstoftwo
 \else \expandafter \@secondoftwo
 \fi
}%
\providecommand \@ifx [1]{%
 \ifx #1\expandafter \@firstoftwo
 \else \expandafter \@secondoftwo
 \fi
}%
\providecommand \natexlab [1]{#1}%
\providecommand \enquote  [1]{``#1''}%
\providecommand \bibnamefont  [1]{#1}%
\providecommand \bibfnamefont [1]{#1}%
\providecommand \citenamefont [1]{#1}%
\providecommand \href@noop [0]{\@secondoftwo}%
\providecommand \href [0]{\begingroup \@sanitize@url \@href}%
\providecommand \@href[1]{\@@startlink{#1}\@@href}%
\providecommand \@@href[1]{\endgroup#1\@@endlink}%
\providecommand \@sanitize@url [0]{\catcode `\\12\catcode `\$12\catcode
  `\&12\catcode `\#12\catcode `\^12\catcode `\_12\catcode `\%12\relax}%
\providecommand \@@startlink[1]{}%
\providecommand \@@endlink[0]{}%
\providecommand \url  [0]{\begingroup\@sanitize@url \@url }%
\providecommand \@url [1]{\endgroup\@href {#1}{\urlprefix }}%
\providecommand \urlprefix  [0]{URL }%
\providecommand \Eprint [0]{\href }%
\providecommand \doibase [0]{http://dx.doi.org/}%
\providecommand \selectlanguage [0]{\@gobble}%
\providecommand \bibinfo  [0]{\@secondoftwo}%
\providecommand \bibfield  [0]{\@secondoftwo}%
\providecommand \translation [1]{[#1]}%
\providecommand \BibitemOpen [0]{}%
\providecommand \bibitemStop [0]{}%
\providecommand \bibitemNoStop [0]{.\EOS\space}%
\providecommand \EOS [0]{\spacefactor3000\relax}%
\providecommand \BibitemShut  [1]{\csname bibitem#1\endcsname}%
\let\auto@bib@innerbib\@empty
\bibitem [{\citenamefont {Braginsky}\ and\ \citenamefont
  {Khalili}(1992)}]{Braginsky}%
  \BibitemOpen
  \bibfield  {author} {\bibinfo {author} {\bibfnamefont {V.~B.}\ \bibnamefont
  {Braginsky}}\ and\ \bibinfo {author} {\bibfnamefont {F.~Y.}\ \bibnamefont
  {Khalili}},\ }\href {http://ebooks.cambridge.org/ref/id/CBO9780511622748}
  {\emph {\bibinfo {title} {{Quantum Measurement}}}},\ edited by\ \bibinfo
  {editor} {\bibfnamefont {K.~S.}\ \bibnamefont {Thorne}}\ (\bibinfo
  {publisher} {Cambridge University Press},\ \bibinfo {address} {Cambridge},\
  \bibinfo {year} {1992})\BibitemShut {NoStop}%
\bibitem [{\citenamefont {Caves}(1980)}]{Caves1980}%
  \BibitemOpen
  \bibfield  {author} {\bibinfo {author} {\bibfnamefont {C.~M.}\ \bibnamefont
  {Caves}},\ }\href {https://link.aps.org/doi/10.1103/PhysRevLett.45.75}
  {\bibfield  {journal} {\bibinfo  {journal} {Phys. Rev. Lett.}\ }\textbf
  {\bibinfo {volume} {45}},\ \bibinfo {pages} {75} (\bibinfo {year}
  {1980})}\BibitemShut {NoStop}%
\bibitem [{\citenamefont {Gardiner}\ and\ \citenamefont
  {Zoller}(2004)}]{Gardiner1991}%
  \BibitemOpen
  \bibfield  {author} {\bibinfo {author} {\bibfnamefont {C.~W.}\ \bibnamefont
  {Gardiner}}\ and\ \bibinfo {author} {\bibfnamefont {P.}~\bibnamefont
  {Zoller}},\ }\href {https://www.springer.com/gp/book/9783540223016} {\emph
  {\bibinfo {title} {{Quantum Noise: A Handbook of Markovian and Non-Markovian
  Quantum Stochastic Methods with Applications to Quantum Optics}}}}\ (\bibinfo
   {publisher} {Springer},\ \bibinfo {year} {2004})\ p.\ \bibinfo {pages}
  {450}\BibitemShut {NoStop}%
\bibitem [{\citenamefont {Clerk}\ \emph {et~al.}(2010)\citenamefont {Clerk},
  \citenamefont {Devoret}, \citenamefont {Girvin}, \citenamefont {Marquardt},\
  and\ \citenamefont {Schoelkopf}}]{Clerk2010a}%
  \BibitemOpen
  \bibfield  {author} {\bibinfo {author} {\bibfnamefont {A.~A.}\ \bibnamefont
  {Clerk}}, \bibinfo {author} {\bibfnamefont {M.~H.}\ \bibnamefont {Devoret}},
  \bibinfo {author} {\bibfnamefont {S.~M.}\ \bibnamefont {Girvin}}, \bibinfo
  {author} {\bibfnamefont {F.}~\bibnamefont {Marquardt}}, \ and\ \bibinfo
  {author} {\bibfnamefont {R.~J.}\ \bibnamefont {Schoelkopf}},\ }\href
  {https://doi.org/10.1103/RevModPhys.82.1155} {\bibfield  {journal} {\bibinfo
  {journal} {Rev. Mod. Phys.}\ }\textbf {\bibinfo {volume} {82}},\ \bibinfo
  {pages} {1155} (\bibinfo {year} {2010})}\BibitemShut {NoStop}%
\bibitem [{\citenamefont {Adhikari}(2014)}]{Adhikari2014}%
  \BibitemOpen
  \bibfield  {author} {\bibinfo {author} {\bibfnamefont {R.~X.}\ \bibnamefont
  {Adhikari}},\ }\href {https://link.aps.org/doi/10.1103/RevModPhys.86.121}
  {\bibfield  {journal} {\bibinfo  {journal} {Rev. Mod. Phys.}\ }\textbf
  {\bibinfo {volume} {86}},\ \bibinfo {pages} {121} (\bibinfo {year}
  {2014})}\BibitemShut {NoStop}%
\bibitem [{\citenamefont {Miao}\ \emph {et~al.}(2018)\citenamefont {Miao},
  \citenamefont {Yang},\ and\ \citenamefont {Martynov}}]{Miao}%
  \BibitemOpen
  \bibfield  {author} {\bibinfo {author} {\bibfnamefont {H.}~\bibnamefont
  {Miao}}, \bibinfo {author} {\bibfnamefont {H.}~\bibnamefont {Yang}}, \ and\
  \bibinfo {author} {\bibfnamefont {D.}~\bibnamefont {Martynov}},\ }\href
  {https://link.aps.org/doi/10.1103/PhysRevD.98.044044} {\bibfield  {journal}
  {\bibinfo  {journal} {Phys. Rev. D}\ }\textbf {\bibinfo {volume} {98}},\
  \bibinfo {pages} {044044} (\bibinfo {year} {2018})}\BibitemShut {NoStop}%
\bibitem [{\citenamefont {Chen}(2013)}]{Chen2013}%
  \BibitemOpen
  \bibfield  {author} {\bibinfo {author} {\bibfnamefont {Y.}~\bibnamefont
  {Chen}},\ }\href {https://doi.org/10.1088/0953} {\bibfield  {journal}
  {\bibinfo  {journal} {J. Phys. B: At. Mol. Opt. Phys.}\ }\textbf {\bibinfo
  {volume} {46}},\ \bibinfo {pages} {104001} (\bibinfo {year}
  {2013})}\BibitemShut {NoStop}%
\bibitem [{\citenamefont {Aspelmeyer}\ \emph {et~al.}(2014)\citenamefont
  {Aspelmeyer}, \citenamefont {Kippenberg},\ and\ \citenamefont
  {Marquardt}}]{Aspelmeyer2014}%
  \BibitemOpen
  \bibfield  {author} {\bibinfo {author} {\bibfnamefont {M.}~\bibnamefont
  {Aspelmeyer}}, \bibinfo {author} {\bibfnamefont {T.~J.}\ \bibnamefont
  {Kippenberg}}, \ and\ \bibinfo {author} {\bibfnamefont {F.}~\bibnamefont
  {Marquardt}},\ }\href {https://doi.org/10.1103/RevModPhys.86.1391} {\bibfield
   {journal} {\bibinfo  {journal} {Rev. Mod. Phys.}\ }\textbf {\bibinfo
  {volume} {86}},\ \bibinfo {pages} {1391} (\bibinfo {year}
  {2014})}\BibitemShut {NoStop}%
\bibitem [{\citenamefont {Sikivie}(1983)}]{Sikivie1983}%
  \BibitemOpen
  \bibfield  {author} {\bibinfo {author} {\bibfnamefont {P.}~\bibnamefont
  {Sikivie}},\ }\href {https://doi.org/10.1103/PhysRevLett.51.1415} {\bibfield
  {journal} {\bibinfo  {journal} {Phys. Rev. Lett.}\ }\textbf {\bibinfo
  {volume} {51}},\ \bibinfo {pages} {1415} (\bibinfo {year}
  {1983})}\BibitemShut {NoStop}%
\bibitem [{\citenamefont {DeRocco}\ and\ \citenamefont
  {Hook}(2018)}]{Derocco2018}%
  \BibitemOpen
  \bibfield  {author} {\bibinfo {author} {\bibfnamefont {W.}~\bibnamefont
  {DeRocco}}\ and\ \bibinfo {author} {\bibfnamefont {A.}~\bibnamefont {Hook}},\
  }\href {https://doi.org/10.1103/PhysRevD.98.035021} {\bibfield  {journal}
  {\bibinfo  {journal} {Phys. Rev. D}\ }\textbf {\bibinfo {volume} {98}},\
  \bibinfo {pages} {35021} (\bibinfo {year} {2018})}\BibitemShut {NoStop}%
\bibitem [{\citenamefont {Abbott}\ \emph {et~al.}(2019)\citenamefont {Abbott},
  \citenamefont {Abbott}, \citenamefont {Abbott} \emph
  {et~al.}}]{TheLIGOScientificCollaboration2019}%
  \BibitemOpen
  \bibfield  {author} {\bibinfo {author} {\bibfnamefont {B.~P.}\ \bibnamefont
  {Abbott}}, \bibinfo {author} {\bibfnamefont {R.}~\bibnamefont {Abbott}},
  \bibinfo {author} {\bibfnamefont {T.~D.}\ \bibnamefont {Abbott}},  \emph
  {et~al.},\ }\href {https://link.aps.org/doi/10.1103/PhysRevX.9.031040}
  {\bibfield  {journal} {\bibinfo  {journal} {Physical Review X}\ }\textbf
  {\bibinfo {volume} {9}},\ \bibinfo {pages} {031040} (\bibinfo {year}
  {2019})}\BibitemShut {NoStop}%
\bibitem [{\citenamefont {James}\ \emph {et~al.}(2008)\citenamefont {James},
  \citenamefont {Nurdin},\ and\ \citenamefont {Petersen}}]{James2008}%
  \BibitemOpen
  \bibfield  {author} {\bibinfo {author} {\bibfnamefont {M.~R.}\ \bibnamefont
  {James}}, \bibinfo {author} {\bibfnamefont {H.~I.}\ \bibnamefont {Nurdin}}, \
  and\ \bibinfo {author} {\bibfnamefont {I.~R.}\ \bibnamefont {Petersen}},\
  }\href {https://ieeexplore.ieee.org/document/4625217} {\bibfield  {journal}
  {\bibinfo  {journal} {IEEE T. Automat. Contr.}\ }\textbf {\bibinfo {volume}
  {53}},\ \bibinfo {pages} {1787} (\bibinfo {year} {2008})}\BibitemShut
  {NoStop}%
\bibitem [{\citenamefont {Mabuchi}(2008)}]{Mabuchi2008a}%
  \BibitemOpen
  \bibfield  {author} {\bibinfo {author} {\bibfnamefont {H.}~\bibnamefont
  {Mabuchi}},\ }\href {https://link.aps.org/doi/10.1103/PhysRevA.78.032323}
  {\bibfield  {journal} {\bibinfo  {journal} {Phys. Rev. A}\ }\textbf {\bibinfo
  {volume} {78}},\ \bibinfo {pages} {032323} (\bibinfo {year}
  {2008})}\BibitemShut {NoStop}%
\bibitem [{\citenamefont {Hamerly}\ and\ \citenamefont
  {Mabuchi}(2012)}]{Hamerly2012}%
  \BibitemOpen
  \bibfield  {author} {\bibinfo {author} {\bibfnamefont {R.}~\bibnamefont
  {Hamerly}}\ and\ \bibinfo {author} {\bibfnamefont {H.}~\bibnamefont
  {Mabuchi}},\ }\href {https://doi.org/10.1103/PhysRevLett.109.173602}
  {\bibfield  {journal} {\bibinfo  {journal} {Phys. Rev. Lett.}\ }\textbf
  {\bibinfo {volume} {109}},\ \bibinfo {pages} {173602} (\bibinfo {year}
  {2012})}\BibitemShut {NoStop}%
\bibitem [{\citenamefont {Jacobs}\ \emph {et~al.}(2014)\citenamefont {Jacobs},
  \citenamefont {Wang},\ and\ \citenamefont {Wiseman}}]{Jacobs2014}%
  \BibitemOpen
  \bibfield  {author} {\bibinfo {author} {\bibfnamefont {K.}~\bibnamefont
  {Jacobs}}, \bibinfo {author} {\bibfnamefont {X.}~\bibnamefont {Wang}}, \ and\
  \bibinfo {author} {\bibfnamefont {H.~M.}\ \bibnamefont {Wiseman}},\ }\href
  {https://doi.org/10.1088/1367-2630/16/7/073036} {\bibfield  {journal}
  {\bibinfo  {journal} {New Journal of Physics}\ }\textbf {\bibinfo {volume}
  {16}} (\bibinfo {year} {2014})}\BibitemShut {NoStop}%
\bibitem [{\citenamefont {Kimble}\ \emph {et~al.}(2001)\citenamefont {Kimble},
  \citenamefont {Levin}, \citenamefont {Matsko}, \citenamefont {Thorne},\ and\
  \citenamefont {Vyatchanin}}]{Kimble2000}%
  \BibitemOpen
  \bibfield  {author} {\bibinfo {author} {\bibfnamefont {H.~J.}\ \bibnamefont
  {Kimble}}, \bibinfo {author} {\bibfnamefont {Y.}~\bibnamefont {Levin}},
  \bibinfo {author} {\bibfnamefont {A.~B.}\ \bibnamefont {Matsko}}, \bibinfo
  {author} {\bibfnamefont {K.~S.}\ \bibnamefont {Thorne}}, \ and\ \bibinfo
  {author} {\bibfnamefont {S.~P.}\ \bibnamefont {Vyatchanin}},\ }\href
  {http://dx.doi.org/10.1103/PhysRevD.65.022002} {\bibfield  {journal}
  {\bibinfo  {journal} {Phys. Rev. D}\ }\textbf {\bibinfo {volume} {65}},\
  \bibinfo {pages} {022002} (\bibinfo {year} {2001})}\BibitemShut {NoStop}%
\bibitem [{\citenamefont {Oelker}\ \emph
  {et~al.}(2016{\natexlab{a}})\citenamefont {Oelker}, \citenamefont {Isogai},
  \citenamefont {Miller}, \citenamefont {Tse}, \citenamefont {Barsotti},
  \citenamefont {Mavalvala},\ and\ \citenamefont {Evans}}]{Oelker2016c}%
  \BibitemOpen
  \bibfield  {author} {\bibinfo {author} {\bibfnamefont {E.}~\bibnamefont
  {Oelker}}, \bibinfo {author} {\bibfnamefont {T.}~\bibnamefont {Isogai}},
  \bibinfo {author} {\bibfnamefont {J.}~\bibnamefont {Miller}}, \bibinfo
  {author} {\bibfnamefont {M.}~\bibnamefont {Tse}}, \bibinfo {author}
  {\bibfnamefont {L.}~\bibnamefont {Barsotti}}, \bibinfo {author}
  {\bibfnamefont {N.}~\bibnamefont {Mavalvala}}, \ and\ \bibinfo {author}
  {\bibfnamefont {M.}~\bibnamefont {Evans}},\ }\href
  {https://link.aps.org/doi/10.1103/PhysRevLett.116.041102} {\bibfield
  {journal} {\bibinfo  {journal} {Phys. Rev. Lett.}\ }\textbf {\bibinfo
  {volume} {116}},\ \bibinfo {pages} {041102} (\bibinfo {year}
  {2016}{\natexlab{a}})}\BibitemShut {NoStop}%
\bibitem [{\citenamefont {Schnabel}(2017)}]{Schnabel2017}%
  \BibitemOpen
  \bibfield  {author} {\bibinfo {author} {\bibfnamefont {R.}~\bibnamefont
  {Schnabel}},\ }\href {http://dx.doi.org/10.1016/j.physrep.2017.04.001}
  {\enquote {\bibinfo {title} {{Squeezed states of light and their applications
  in laser interferometers}},}\ } (\bibinfo {year} {2017})\BibitemShut
  {NoStop}%
\bibitem [{\citenamefont {Gough}\ and\ \citenamefont
  {James}(2009{\natexlab{a}})}]{Gough2007}%
  \BibitemOpen
  \bibfield  {author} {\bibinfo {author} {\bibfnamefont {J.}~\bibnamefont
  {Gough}}\ and\ \bibinfo {author} {\bibfnamefont {M.~R.}\ \bibnamefont
  {James}},\ }\href {http://arxiv.org/abs/0707.0048} {\bibfield  {journal}
  {\bibinfo  {journal} {IEEE T. Automat. Contr.}\ }\textbf {\bibinfo {volume}
  {54}},\ \bibinfo {pages} {2530} (\bibinfo {year}
  {2009}{\natexlab{a}})}\BibitemShut {NoStop}%
\bibitem [{\citenamefont {Gough}\ \emph {et~al.}(2010)\citenamefont {Gough},
  \citenamefont {James},\ and\ \citenamefont {Nurdin}}]{Gough2009}%
  \BibitemOpen
  \bibfield  {author} {\bibinfo {author} {\bibfnamefont {J.~E.}\ \bibnamefont
  {Gough}}, \bibinfo {author} {\bibfnamefont {M.~R.}\ \bibnamefont {James}}, \
  and\ \bibinfo {author} {\bibfnamefont {H.~I.}\ \bibnamefont {Nurdin}},\
  }\href {http://dx.doi.org/10.1103/PhysRevA.81.023804} {\bibfield  {journal}
  {\bibinfo  {journal} {Phys. Rev. A}\ }\textbf {\bibinfo {volume} {81}},\
  \bibinfo {pages} {023804} (\bibinfo {year} {2010})}\BibitemShut {NoStop}%
\bibitem [{\citenamefont {Gough}\ and\ \citenamefont
  {James}(2009{\natexlab{b}})}]{Gough2009a}%
  \BibitemOpen
  \bibfield  {author} {\bibinfo {author} {\bibfnamefont {J.}~\bibnamefont
  {Gough}}\ and\ \bibinfo {author} {\bibfnamefont {M.~R.}\ \bibnamefont
  {James}},\ }\href {https://doi.org/10.1007/s00220-008-0698-8} {\bibfield
  {journal} {\bibinfo  {journal} {Comm. Math. Phys.}\ }\textbf {\bibinfo
  {volume} {287}},\ \bibinfo {pages} {1109} (\bibinfo {year}
  {2009}{\natexlab{b}})}\BibitemShut {NoStop}%
\bibitem [{\citenamefont {Tezak}\ \emph {et~al.}(2012)\citenamefont {Tezak},
  \citenamefont {Niederberger}, \citenamefont {Pavlichin}, \citenamefont
  {Sarma},\ and\ \citenamefont {Mabuchi}}]{Tezak2012a}%
  \BibitemOpen
  \bibfield  {author} {\bibinfo {author} {\bibfnamefont {N.}~\bibnamefont
  {Tezak}}, \bibinfo {author} {\bibfnamefont {A.}~\bibnamefont {Niederberger}},
  \bibinfo {author} {\bibfnamefont {D.~S.}\ \bibnamefont {Pavlichin}}, \bibinfo
  {author} {\bibfnamefont {G.}~\bibnamefont {Sarma}}, \ and\ \bibinfo {author}
  {\bibfnamefont {H.}~\bibnamefont {Mabuchi}},\ }\href
  {https://doi.org/10.1098/rsta.2011.0526} {\bibfield  {journal} {\bibinfo
  {journal} {Philos. T. R. Soc. A}\ }\textbf {\bibinfo {volume} {370}},\
  \bibinfo {pages} {5270} (\bibinfo {year} {2012})}\BibitemShut {NoStop}%
\bibitem [{\citenamefont {Combes}\ \emph {et~al.}(2017)\citenamefont {Combes},
  \citenamefont {Kerckhoff},\ and\ \citenamefont {Sarovar}}]{Combes2016}%
  \BibitemOpen
  \bibfield  {author} {\bibinfo {author} {\bibfnamefont {J.}~\bibnamefont
  {Combes}}, \bibinfo {author} {\bibfnamefont {J.}~\bibnamefont {Kerckhoff}}, \
  and\ \bibinfo {author} {\bibfnamefont {M.}~\bibnamefont {Sarovar}},\ }\href
  {http://dx.doi.org/10.1080/23746149.2017.1343097} {\bibfield  {journal}
  {\bibinfo  {journal} {Advances in Physics: X}\ }\textbf {\bibinfo {volume}
  {2}},\ \bibinfo {pages} {784} (\bibinfo {year} {2017})}\BibitemShut {NoStop}%
\bibitem [{\citenamefont {Nurdin}\ \emph {et~al.}(2009)\citenamefont {Nurdin},
  \citenamefont {James},\ and\ \citenamefont {Doherty}}]{Nurdin2009}%
  \BibitemOpen
  \bibfield  {author} {\bibinfo {author} {\bibfnamefont {H.~I.}\ \bibnamefont
  {Nurdin}}, \bibinfo {author} {\bibfnamefont {M.~R.}\ \bibnamefont {James}}, \
  and\ \bibinfo {author} {\bibfnamefont {A.~C.}\ \bibnamefont {Doherty}},\
  }\href {http://epubs.siam.org/doi/10.1137/080728652} {\bibfield  {journal}
  {\bibinfo  {journal} {SIAM J. Control. Optim.}\ }\textbf {\bibinfo {volume}
  {48}},\ \bibinfo {pages} {2686} (\bibinfo {year} {2009})}\BibitemShut
  {NoStop}%
\bibitem [{\citenamefont {Nurdin}(2010{\natexlab{a}})}]{Nurdin2010a}%
  \BibitemOpen
  \bibfield  {author} {\bibinfo {author} {\bibfnamefont {H.~I.}\ \bibnamefont
  {Nurdin}},\ }\href {https://doi.org/10.1109/TAC.2010.2062892} {\bibfield
  {journal} {\bibinfo  {journal} {IEEE T. Automat. Contr.}\ }\textbf {\bibinfo
  {volume} {55}},\ \bibinfo {pages} {2439} (\bibinfo {year}
  {2010}{\natexlab{a}})}\BibitemShut {NoStop}%
\bibitem [{\citenamefont {Nurdin}(2010{\natexlab{b}})}]{Nurdin2010b}%
  \BibitemOpen
  \bibfield  {author} {\bibinfo {author} {\bibfnamefont {H.}~\bibnamefont
  {Nurdin}},\ }\href {http://ieeexplore.ieee.org/document/5404766/} {\bibfield
  {journal} {\bibinfo  {journal} {IEEE T. Automat. Contr.}\ }\textbf {\bibinfo
  {volume} {55}},\ \bibinfo {pages} {1008} (\bibinfo {year}
  {2010}{\natexlab{b}})}\BibitemShut {NoStop}%
\bibitem [{\citenamefont {Nurdin}\ \emph {et~al.}(2016)\citenamefont {Nurdin},
  \citenamefont {Grivopoulos},\ and\ \citenamefont {Petersen}}]{Nurdin2016}%
  \BibitemOpen
  \bibfield  {author} {\bibinfo {author} {\bibfnamefont {H.~I.}\ \bibnamefont
  {Nurdin}}, \bibinfo {author} {\bibfnamefont {S.}~\bibnamefont {Grivopoulos}},
  \ and\ \bibinfo {author} {\bibfnamefont {I.~R.}\ \bibnamefont {Petersen}},\
  }\href {https://doi.org/10.1016/j.automatica.2016.03.002} {\bibfield
  {journal} {\bibinfo  {journal} {Automatica}\ }\textbf {\bibinfo {volume}
  {69}},\ \bibinfo {pages} {324} (\bibinfo {year} {2016})}\BibitemShut
  {NoStop}%
\bibitem [{\citenamefont {Grivopoulos}\ \emph {et~al.}(2016)\citenamefont
  {Grivopoulos}, \citenamefont {Nurdin},\ and\ \citenamefont
  {Petersen}}]{Grivopoulos2016}%
  \BibitemOpen
  \bibfield  {author} {\bibinfo {author} {\bibfnamefont {S.}~\bibnamefont
  {Grivopoulos}}, \bibinfo {author} {\bibfnamefont {H.~I.}\ \bibnamefont
  {Nurdin}}, \ and\ \bibinfo {author} {\bibfnamefont {I.~R.}\ \bibnamefont
  {Petersen}},\ }\href {https://doi.org/10.1109/CDC.2016.7798962} {\bibfield
  {journal} {\bibinfo  {journal} {IEEE Decis. Contr. P.}\ }\textbf {\bibinfo
  {volume} {110100020}},\ \bibinfo {pages} {4552} (\bibinfo {year}
  {2016})}\BibitemShut {NoStop}%
\bibitem [{\citenamefont {Nurdin}\ and\ \citenamefont
  {Yamamoto}(2017)}]{Nurdin2017}%
  \BibitemOpen
  \bibfield  {author} {\bibinfo {author} {\bibfnamefont {H.~I.}\ \bibnamefont
  {Nurdin}}\ and\ \bibinfo {author} {\bibfnamefont {N.}~\bibnamefont
  {Yamamoto}},\ }\href {http://link.springer.com/10.1007/978-3-319-55201-9}
  {\emph {\bibinfo {title} {{Linear Dynamical Quantum Systems: Analysis,
  Synthesis, and Control}}}}\ (\bibinfo  {publisher} {Springer},\ \bibinfo
  {year} {2017})\BibitemShut {NoStop}%
\bibitem [{\citenamefont {Grivopoulos}\ and\ \citenamefont
  {Petersen}(2017)}]{Grivopoulos2017}%
  \BibitemOpen
  \bibfield  {author} {\bibinfo {author} {\bibfnamefont {S.}~\bibnamefont
  {Grivopoulos}}\ and\ \bibinfo {author} {\bibfnamefont {I.}~\bibnamefont
  {Petersen}},\ }\href {https://epubs.siam.org/doi/10.1137/15M104829X}
  {\bibfield  {journal} {\bibinfo  {journal} {SIAM J. Control. Optim.}\
  }\textbf {\bibinfo {volume} {55}},\ \bibinfo {pages} {3349} (\bibinfo {year}
  {2017})}\BibitemShut {NoStop}%
\bibitem [{\citenamefont {Petersen}\ \emph {et~al.}(2018)\citenamefont
  {Petersen}, \citenamefont {James}, \citenamefont {Ugrinovskii},\ and\
  \citenamefont {Yamamoto}}]{Petersen2018}%
  \BibitemOpen
  \bibfield  {author} {\bibinfo {author} {\bibfnamefont {I.~R.}\ \bibnamefont
  {Petersen}}, \bibinfo {author} {\bibfnamefont {M.~R.}\ \bibnamefont {James}},
  \bibinfo {author} {\bibfnamefont {V.}~\bibnamefont {Ugrinovskii}}, \ and\
  \bibinfo {author} {\bibfnamefont {N.}~\bibnamefont {Yamamoto}},\ }in\ \href
  {http://arxiv.org/abs/1802.03887} {\emph {\bibinfo {booktitle} {E. C. C.}}}\
  (\bibinfo {year} {2018})\ pp.\ \bibinfo {pages} {3185--3190}\BibitemShut
  {NoStop}%
\bibitem [{\citenamefont {Shaiju}\ and\ \citenamefont
  {Petersen}(2012)}]{Shaiju2012}%
  \BibitemOpen
  \bibfield  {author} {\bibinfo {author} {\bibfnamefont {A.~J.}\ \bibnamefont
  {Shaiju}}\ and\ \bibinfo {author} {\bibfnamefont {I.~R.}\ \bibnamefont
  {Petersen}},\ }\href {https://doi.org/10.1109/TAC.2012.2195929} {\bibfield
  {journal} {\bibinfo  {journal} {IEEE T. Automat. Contr.}\ }\textbf {\bibinfo
  {volume} {57}},\ \bibinfo {pages} {2033} (\bibinfo {year}
  {2012})}\BibitemShut {NoStop}%
\bibitem [{\citenamefont {Luenberger}(1967)}]{Luenberger1967}%
  \BibitemOpen
  \bibfield  {author} {\bibinfo {author} {\bibfnamefont {D.}~\bibnamefont
  {Luenberger}},\ }\href {http://ieeexplore.ieee.org/document/1098584/}
  {\bibfield  {journal} {\bibinfo  {journal} {IEEE T. Automat. Contr.}\
  }\textbf {\bibinfo {volume} {12}},\ \bibinfo {pages} {290} (\bibinfo {year}
  {1967})}\BibitemShut {NoStop}%
\bibitem [{\citenamefont {Ackermann}\ and\ \citenamefont
  {Bucy}(1971)}]{Ackermann1971}%
  \BibitemOpen
  \bibfield  {author} {\bibinfo {author} {\bibfnamefont {J.}~\bibnamefont
  {Ackermann}}\ and\ \bibinfo {author} {\bibfnamefont {R.}~\bibnamefont
  {Bucy}},\ }\href
  {https://linkinghub.elsevier.com/retrieve/pii/S0019995871901057} {\bibfield
  {journal} {\bibinfo  {journal} {Inform. Control}\ }\textbf {\bibinfo {volume}
  {19}},\ \bibinfo {pages} {224} (\bibinfo {year} {1971})}\BibitemShut
  {NoStop}%
\bibitem [{\citenamefont {Kailath}(1980)}]{Kailath1980}%
  \BibitemOpen
  \bibfield  {author} {\bibinfo {author} {\bibfnamefont {T.}~\bibnamefont
  {Kailath}},\ }\href
  {https://books.google.co.uk/books/about/Linear_Systems.html?id=ggYqAQAAMAAJ}
  {\emph {\bibinfo {title} {{Linear Systems}}}},\ \bibinfo {edition} {1st}\
  ed.\ (\bibinfo  {publisher} {Prentice-Hall, Inc.},\ \bibinfo {year} {1980})\
  p.~\bibinfo {pages} {31}\BibitemShut {NoStop}%
\bibitem [{\citenamefont {Antoniou}\ \emph {et~al.}(1988)\citenamefont
  {Antoniou}, \citenamefont {Paraskevopoulos},\ and\ \citenamefont
  {Varoufakis}}]{Antoniou1988}%
  \BibitemOpen
  \bibfield  {author} {\bibinfo {author} {\bibfnamefont {G.~E.}\ \bibnamefont
  {Antoniou}}, \bibinfo {author} {\bibfnamefont {P.~N.}\ \bibnamefont
  {Paraskevopoulos}}, \ and\ \bibinfo {author} {\bibfnamefont {S.~J.}\
  \bibnamefont {Varoufakis}},\ }\href {https://doi.org/10.1109/31.1857}
  {\bibfield  {journal} {\bibinfo  {journal} {IEEE T. Circuits Syst.}\ }\textbf
  {\bibinfo {volume} {35}},\ \bibinfo {pages} {1055} (\bibinfo {year}
  {1988})}\BibitemShut {NoStop}%
\bibitem [{Note5()}]{Note5}%
  \BibitemOpen
  \bibinfo {note} {An excellent primer to state-space representations of
  dynamical systems can be found in~\cite {Bechhoefer2005}.}\BibitemShut
  {Stop}%
\bibitem [{\citenamefont {Blow}\ \emph {et~al.}(1990)\citenamefont {Blow},
  \citenamefont {Loudon}, \citenamefont {Phoenix},\ and\ \citenamefont
  {Shepherd}}]{Blow1990}%
  \BibitemOpen
  \bibfield  {author} {\bibinfo {author} {\bibfnamefont {K.~J.}\ \bibnamefont
  {Blow}}, \bibinfo {author} {\bibfnamefont {R.}~\bibnamefont {Loudon}},
  \bibinfo {author} {\bibfnamefont {S.~J.~D.}\ \bibnamefont {Phoenix}}, \ and\
  \bibinfo {author} {\bibfnamefont {T.~J.}\ \bibnamefont {Shepherd}},\ }\href
  {https://link.aps.org/doi/10.1103/PhysRevA.42.4102} {\bibfield  {journal}
  {\bibinfo  {journal} {Phys. Rev. A}\ }\textbf {\bibinfo {volume} {42}},\
  \bibinfo {pages} {4102} (\bibinfo {year} {1990})}\BibitemShut {NoStop}%
\bibitem [{\citenamefont {Hudson}\ and\ \citenamefont
  {Parthasarathy}(1984)}]{Hudson1984}%
  \BibitemOpen
  \bibfield  {author} {\bibinfo {author} {\bibfnamefont {R.~L.}\ \bibnamefont
  {Hudson}}\ and\ \bibinfo {author} {\bibfnamefont {K.~R.}\ \bibnamefont
  {Parthasarathy}},\ }\href
  {https://projecteuclid.org/download/pdf_1/euclid.cmp/1103941122} {\bibfield
  {journal} {\bibinfo  {journal} {Comm. Math. Phys.}\ }\textbf {\bibinfo
  {volume} {93}},\ \bibinfo {pages} {301} (\bibinfo {year} {1984})}\BibitemShut
  {NoStop}%
\bibitem [{\citenamefont {Parthasarathy}(1992)}]{Parthasarathy1992}%
  \BibitemOpen
  \bibfield  {author} {\bibinfo {author} {\bibfnamefont {K.~R.}\ \bibnamefont
  {Parthasarathy}},\ }\href
  {http://link.springer.com/10.1007/978-3-0348-8641-3} {\emph {\bibinfo {title}
  {{An Introduction to Quantum Stochastic Calculus}}}}\ (\bibinfo  {publisher}
  {Birkh{\"{a}}user Basel},\ \bibinfo {address} {Basel},\ \bibinfo {year}
  {1992})\BibitemShut {NoStop}%
\bibitem [{\citenamefont {Bouten}\ \emph {et~al.}(2007)\citenamefont {Bouten},
  \citenamefont {{Van Handel}},\ and\ \citenamefont {James}}]{Bouten2006}%
  \BibitemOpen
  \bibfield  {author} {\bibinfo {author} {\bibfnamefont {L.}~\bibnamefont
  {Bouten}}, \bibinfo {author} {\bibfnamefont {R.}~\bibnamefont {{Van
  Handel}}}, \ and\ \bibinfo {author} {\bibfnamefont {M.~R.}\ \bibnamefont
  {James}},\ }\href {http://arxiv.org/abs/math/0601741} {\bibfield  {journal}
  {\bibinfo  {journal} {SIAM J. Control. Optim.}\ }\textbf {\bibinfo {volume}
  {46}},\ \bibinfo {pages} {2199} (\bibinfo {year} {2007})}\BibitemShut
  {NoStop}%
\bibitem [{Note1()}]{Note1}%
  \BibitemOpen
  \bibinfo {note} {As discussed in Section.~\ref {appendix:r-matrix} of this
  paper, this matrix takes a different form when using Hermitian observable
  quadrature operators.}\BibitemShut {Stop}%
\bibitem [{Note2()}]{Note2}%
  \BibitemOpen
  \bibinfo {note} {Note that the approach is also entirely general to
  optomechanical systems, provided that the dynamics can be
  linearised.}\BibitemShut {Stop}%
\bibitem [{\citenamefont {Wicht}\ \emph {et~al.}(1997)\citenamefont {Wicht},
  \citenamefont {Danzmann}, \citenamefont {Fleischhauer}, \citenamefont
  {Scully}, \citenamefont {M{\"{u}}ller},\ and\ \citenamefont
  {Rinkleff}}]{Wicht1997}%
  \BibitemOpen
  \bibfield  {author} {\bibinfo {author} {\bibfnamefont {A.}~\bibnamefont
  {Wicht}}, \bibinfo {author} {\bibfnamefont {K.}~\bibnamefont {Danzmann}},
  \bibinfo {author} {\bibfnamefont {M.}~\bibnamefont {Fleischhauer}}, \bibinfo
  {author} {\bibfnamefont {M.}~\bibnamefont {Scully}}, \bibinfo {author}
  {\bibfnamefont {G.}~\bibnamefont {M{\"{u}}ller}}, \ and\ \bibinfo {author}
  {\bibfnamefont {R.~H.}\ \bibnamefont {Rinkleff}},\ }\href
  {https://doi.org/10.1016/S0030-4018(96)00579-2} {\bibfield  {journal}
  {\bibinfo  {journal} {Opt. Commun.}\ }\textbf {\bibinfo {volume} {134}},\
  \bibinfo {pages} {431} (\bibinfo {year} {1997})}\BibitemShut {NoStop}%
\bibitem [{\citenamefont {M{\"{u}}ller}\ \emph {et~al.}(2000)\citenamefont
  {M{\"{u}}ller}, \citenamefont {Homann}, \citenamefont {Rinkleff},
  \citenamefont {Wicht},\ and\ \citenamefont {Danzmann}}]{Muller2000}%
  \BibitemOpen
  \bibfield  {author} {\bibinfo {author} {\bibfnamefont {M.}~\bibnamefont
  {M{\"{u}}ller}}, \bibinfo {author} {\bibfnamefont {F.}~\bibnamefont
  {Homann}}, \bibinfo {author} {\bibfnamefont {R.~H.}\ \bibnamefont
  {Rinkleff}}, \bibinfo {author} {\bibfnamefont {A.}~\bibnamefont {Wicht}}, \
  and\ \bibinfo {author} {\bibfnamefont {K.}~\bibnamefont {Danzmann}},\ }\href
  {https://doi.org/10.1103/PhysRevA.62.060501} {\bibfield  {journal} {\bibinfo
  {journal} {Phys. Rev. A}\ }\textbf {\bibinfo {volume} {62}},\ \bibinfo
  {pages} {060501(R)} (\bibinfo {year} {2000})}\BibitemShut {NoStop}%
\bibitem [{\citenamefont {Wise}\ \emph {et~al.}(2004)\citenamefont {Wise},
  \citenamefont {Mueller}, \citenamefont {Reitze}, \citenamefont {Tanner},\
  and\ \citenamefont {Whiting}}]{Wise2004}%
  \BibitemOpen
  \bibfield  {author} {\bibinfo {author} {\bibfnamefont {S.}~\bibnamefont
  {Wise}}, \bibinfo {author} {\bibfnamefont {G.}~\bibnamefont {Mueller}},
  \bibinfo {author} {\bibfnamefont {D.}~\bibnamefont {Reitze}}, \bibinfo
  {author} {\bibfnamefont {D.~B.}\ \bibnamefont {Tanner}}, \ and\ \bibinfo
  {author} {\bibfnamefont {B.~F.}\ \bibnamefont {Whiting}},\ }\href
  {https://doi.org/10.1088/0264-9381/21/5/097} {\bibfield  {journal} {\bibinfo
  {journal} {Class. Quantum Grav.}\ }\textbf {\bibinfo {volume} {21}},\
  \bibinfo {pages} {S1031} (\bibinfo {year} {2004})}\BibitemShut {NoStop}%
\bibitem [{\citenamefont {Pati}\ \emph {et~al.}(2007)\citenamefont {Pati},
  \citenamefont {Salit}, \citenamefont {Salit},\ and\ \citenamefont
  {Shahriar}}]{Pati2007}%
  \BibitemOpen
  \bibfield  {author} {\bibinfo {author} {\bibfnamefont {G.~S.}\ \bibnamefont
  {Pati}}, \bibinfo {author} {\bibfnamefont {M.}~\bibnamefont {Salit}},
  \bibinfo {author} {\bibfnamefont {K.}~\bibnamefont {Salit}}, \ and\ \bibinfo
  {author} {\bibfnamefont {M.~S.}\ \bibnamefont {Shahriar}},\ }\href
  {https://doi.org/10.1103/PhysRevLett.99.133601} {\bibfield  {journal}
  {\bibinfo  {journal} {Phys. Rev. Lett.}\ }\textbf {\bibinfo {volume} {99}},\
  \bibinfo {pages} {133601} (\bibinfo {year} {2007})}\BibitemShut {NoStop}%
\bibitem [{\citenamefont {Yum}\ \emph {et~al.}(2013)\citenamefont {Yum},
  \citenamefont {Scheuer}, \citenamefont {Salit}, \citenamefont {Hemmer},\ and\
  \citenamefont {Shahriar}}]{Yum2013}%
  \BibitemOpen
  \bibfield  {author} {\bibinfo {author} {\bibfnamefont {H.~N.}\ \bibnamefont
  {Yum}}, \bibinfo {author} {\bibfnamefont {J.}~\bibnamefont {Scheuer}},
  \bibinfo {author} {\bibfnamefont {M.}~\bibnamefont {Salit}}, \bibinfo
  {author} {\bibfnamefont {P.~R.}\ \bibnamefont {Hemmer}}, \ and\ \bibinfo
  {author} {\bibfnamefont {M.~S.}\ \bibnamefont {Shahriar}},\ }\href
  {https://www.osapublishing.org/jlt/abstract.cfm?uri=jlt-31-23-3865}
  {\bibfield  {journal} {\bibinfo  {journal} {J. Lightwave Technol.}\ }\textbf
  {\bibinfo {volume} {31}},\ \bibinfo {pages} {3865} (\bibinfo {year}
  {2013})}\BibitemShut {NoStop}%
\bibitem [{\citenamefont {Ma}\ \emph {et~al.}(2015)\citenamefont {Ma},
  \citenamefont {Miao}, \citenamefont {Zhao},\ and\ \citenamefont
  {Chen}}]{Ma2015}%
  \BibitemOpen
  \bibfield  {author} {\bibinfo {author} {\bibfnamefont {Y.}~\bibnamefont
  {Ma}}, \bibinfo {author} {\bibfnamefont {H.}~\bibnamefont {Miao}}, \bibinfo
  {author} {\bibfnamefont {C.}~\bibnamefont {Zhao}}, \ and\ \bibinfo {author}
  {\bibfnamefont {Y.}~\bibnamefont {Chen}},\ }\href
  {https://link.aps.org/doi/10.1103/PhysRevA.92.023807} {\bibfield  {journal}
  {\bibinfo  {journal} {Phys. Rev. A}\ }\textbf {\bibinfo {volume} {92}},\
  \bibinfo {pages} {023807} (\bibinfo {year} {2015})}\BibitemShut {NoStop}%
\bibitem [{\citenamefont {Miao}\ \emph {et~al.}(2015)\citenamefont {Miao},
  \citenamefont {Ma}, \citenamefont {Zhao},\ and\ \citenamefont
  {Chen}}]{Miao2015}%
  \BibitemOpen
  \bibfield  {author} {\bibinfo {author} {\bibfnamefont {H.}~\bibnamefont
  {Miao}}, \bibinfo {author} {\bibfnamefont {Y.}~\bibnamefont {Ma}}, \bibinfo
  {author} {\bibfnamefont {C.}~\bibnamefont {Zhao}}, \ and\ \bibinfo {author}
  {\bibfnamefont {Y.}~\bibnamefont {Chen}},\ }\href
  {https://doi.org/10.1103/PhysRevLett.115.211104} {\bibfield  {journal}
  {\bibinfo  {journal} {Phys. Rev. Lett.}\ }\textbf {\bibinfo {volume} {115}},\
  \bibinfo {pages} {211104} (\bibinfo {year} {2015})}\BibitemShut {NoStop}%
\bibitem [{\citenamefont {David}\ \emph {et~al.}(2015)\citenamefont {David},
  \citenamefont {Li}, \citenamefont {Chunnong} \emph {et~al.}}]{David2015}%
  \BibitemOpen
  \bibfield  {author} {\bibinfo {author} {\bibfnamefont {B.}~\bibnamefont
  {David}}, \bibinfo {author} {\bibfnamefont {J.~U.}\ \bibnamefont {Li}},
  \bibinfo {author} {\bibfnamefont {Z.}~\bibnamefont {Chunnong}},  \emph
  {et~al.},\ }\href
  {http://engine.scichina.com/publisher/scp/journal/SCPMA/58/12/10.1007/s11433-015-5747-7?slug=abstract}
  {\bibfield  {journal} {\bibinfo  {journal} {Sci. China Phys. Mech.}\ }\textbf
  {\bibinfo {volume} {58}},\ \bibinfo {pages} {120405} (\bibinfo {year}
  {2015})}\BibitemShut {NoStop}%
\bibitem [{\citenamefont {Page}\ \emph {et~al.}(2018)\citenamefont {Page},
  \citenamefont {Qin}, \citenamefont {{La Fontaine}}, \citenamefont {Zhao},
  \citenamefont {Ju},\ and\ \citenamefont {Blair}}]{Page2018}%
  \BibitemOpen
  \bibfield  {author} {\bibinfo {author} {\bibfnamefont {M.}~\bibnamefont
  {Page}}, \bibinfo {author} {\bibfnamefont {J.}~\bibnamefont {Qin}}, \bibinfo
  {author} {\bibfnamefont {J.}~\bibnamefont {{La Fontaine}}}, \bibinfo {author}
  {\bibfnamefont {C.}~\bibnamefont {Zhao}}, \bibinfo {author} {\bibfnamefont
  {L.}~\bibnamefont {Ju}}, \ and\ \bibinfo {author} {\bibfnamefont
  {D.}~\bibnamefont {Blair}},\ }\href
  {https://doi.org/10.1103/PhysRevD.97.124060} {\bibfield  {journal} {\bibinfo
  {journal} {Phys. Rev. D}\ }\textbf {\bibinfo {volume} {97}},\ \bibinfo
  {pages} {124060} (\bibinfo {year} {2018})}\BibitemShut {NoStop}%
\bibitem [{\citenamefont {Bentley}\ \emph {et~al.}(2019)\citenamefont
  {Bentley}, \citenamefont {Jones}, \citenamefont {Martynov}, \citenamefont
  {Freise},\ and\ \citenamefont {Miao}}]{Bentley2019}%
  \BibitemOpen
  \bibfield  {author} {\bibinfo {author} {\bibfnamefont {J.}~\bibnamefont
  {Bentley}}, \bibinfo {author} {\bibfnamefont {P.}~\bibnamefont {Jones}},
  \bibinfo {author} {\bibfnamefont {D.}~\bibnamefont {Martynov}}, \bibinfo
  {author} {\bibfnamefont {A.}~\bibnamefont {Freise}}, \ and\ \bibinfo {author}
  {\bibfnamefont {H.}~\bibnamefont {Miao}},\ }\href
  {https://link.aps.org/doi/10.1103/PhysRevD.99.102001} {\bibfield  {journal}
  {\bibinfo  {journal} {Phys. Rev. D}\ }\textbf {\bibinfo {volume} {99}},\
  \bibinfo {pages} {102001} (\bibinfo {year} {2019})}\BibitemShut {NoStop}%
\bibitem [{\citenamefont {Zhou}\ \emph {et~al.}(2015)\citenamefont {Zhou},
  \citenamefont {Zhou},\ and\ \citenamefont {Shahriar}}]{Zhou2015}%
  \BibitemOpen
  \bibfield  {author} {\bibinfo {author} {\bibfnamefont {M.}~\bibnamefont
  {Zhou}}, \bibinfo {author} {\bibfnamefont {Z.}~\bibnamefont {Zhou}}, \ and\
  \bibinfo {author} {\bibfnamefont {S.~M.}\ \bibnamefont {Shahriar}},\ }\href
  {https://link.aps.org/doi/10.1103/PhysRevD.92.082002} {\bibfield  {journal}
  {\bibinfo  {journal} {Phys. Rev. D}\ }\textbf {\bibinfo {volume} {92}},\
  \bibinfo {pages} {082002} (\bibinfo {year} {2015})}\BibitemShut {NoStop}%
\bibitem [{\citenamefont {Zhou}\ and\ \citenamefont
  {Shahriar}(2018)}]{Zhou2018}%
  \BibitemOpen
  \bibfield  {author} {\bibinfo {author} {\bibfnamefont {M.}~\bibnamefont
  {Zhou}}\ and\ \bibinfo {author} {\bibfnamefont {S.~M.}\ \bibnamefont
  {Shahriar}},\ }\href {https://doi.org/10.1103/PhysRevD.98.022003} {\bibfield
  {journal} {\bibinfo  {journal} {Phys. Rev. D}\ }\textbf {\bibinfo {volume}
  {98}},\ \bibinfo {pages} {22003} (\bibinfo {year} {2018})}\BibitemShut
  {NoStop}%
\bibitem [{\citenamefont {Page}\ \emph {et~al.}(2019)\citenamefont {Page},
  \citenamefont {Goryachev}, \citenamefont {Ma}, \citenamefont {Blair},
  \citenamefont {Ju}, \citenamefont {Blair}, \citenamefont {Tobar},\ and\
  \citenamefont {Zhao}}]{Page2019}%
  \BibitemOpen
  \bibfield  {author} {\bibinfo {author} {\bibfnamefont {M.~A.}\ \bibnamefont
  {Page}}, \bibinfo {author} {\bibfnamefont {M.}~\bibnamefont {Goryachev}},
  \bibinfo {author} {\bibfnamefont {Y.}~\bibnamefont {Ma}}, \bibinfo {author}
  {\bibfnamefont {C.~D.}\ \bibnamefont {Blair}}, \bibinfo {author}
  {\bibfnamefont {L.}~\bibnamefont {Ju}}, \bibinfo {author} {\bibfnamefont
  {D.~G.}\ \bibnamefont {Blair}}, \bibinfo {author} {\bibfnamefont {M.~E.}\
  \bibnamefont {Tobar}}, \ and\ \bibinfo {author} {\bibfnamefont
  {C.}~\bibnamefont {Zhao}},\ }\href@noop {} {\bibfield  {journal} {\bibinfo
  {journal} {LIGO DCC}\ } (\bibinfo {year} {2019})}\BibitemShut {NoStop}%
\bibitem [{\citenamefont {Shimazu}\ and\ \citenamefont
  {Yamamoto}(2019)}]{Shimazu2019}%
  \BibitemOpen
  \bibfield  {author} {\bibinfo {author} {\bibfnamefont {R.}~\bibnamefont
  {Shimazu}}\ and\ \bibinfo {author} {\bibfnamefont {N.}~\bibnamefont
  {Yamamoto}},\ }\href {http://arxiv.org/abs/1909.12822} {\bibfield  {journal}
  {\bibinfo  {journal} {arXiv:1909.12822 [quant-ph]}\ } (\bibinfo {year}
  {2019})}\BibitemShut {NoStop}%
\bibitem [{\citenamefont {Mizuno}(1995)}]{JunMizuno}%
  \BibitemOpen
  \bibfield  {author} {\bibinfo {author} {\bibfnamefont {J.}~\bibnamefont
  {Mizuno}},\ }\href
  {http://www2.mpq.mpg.de/$\sim$ros/geo600_docu/text/theses/Jun_Mizuno.pdf}
  {\bibfield  {journal} {\bibinfo  {journal} {Thesis}\ } (\bibinfo {year}
  {1995})}\BibitemShut {NoStop}%
\bibitem [{\citenamefont {{de L. Kronig}}(1926)}]{Kronig1926}%
  \BibitemOpen
  \bibfield  {author} {\bibinfo {author} {\bibfnamefont {R.}~\bibnamefont {{de
  L. Kronig}}},\ }\href {https://link.aps.org/doi/10.1103/PhysRev.49.332}
  {\bibfield  {journal} {\bibinfo  {journal} {J. Opt. Soc. Am.}\ }\textbf
  {\bibinfo {volume} {12}},\ \bibinfo {pages} {547} (\bibinfo {year}
  {1926})}\BibitemShut {NoStop}%
\bibitem [{\citenamefont {Toll}(1956)}]{Toll1956a}%
  \BibitemOpen
  \bibfield  {author} {\bibinfo {author} {\bibfnamefont {J.~S.}\ \bibnamefont
  {Toll}},\ }\href {https://link.aps.org/doi/10.1103/PhysRev.104.1760}
  {\bibfield  {journal} {\bibinfo  {journal} {Physical Review}\ }\textbf
  {\bibinfo {volume} {104}},\ \bibinfo {pages} {1760} (\bibinfo {year}
  {1956})}\BibitemShut {NoStop}%
\bibitem [{\citenamefont {Doyle}\ \emph {et~al.}(1990)\citenamefont {Doyle},
  \citenamefont {Francis},\ and\ \citenamefont {Tannenbaum}}]{Doyle1990}%
  \BibitemOpen
  \bibfield  {author} {\bibinfo {author} {\bibfnamefont {J.}~\bibnamefont
  {Doyle}}, \bibinfo {author} {\bibfnamefont {B.}~\bibnamefont {Francis}}, \
  and\ \bibinfo {author} {\bibfnamefont {A.}~\bibnamefont {Tannenbaum}},\
  }\href {http://link.springer.com/10.1007/978-3-319-07275-3_1} {\emph
  {\bibinfo {title} {{Feedback Control Theory}}}},\ \bibinfo {edition} {1st}\
  ed.\ (\bibinfo  {publisher} {Macmillan Publishing Co.},\ \bibinfo {year}
  {1990})\ Chap.~\bibinfo {chapter} {6}\BibitemShut {NoStop}%
\bibitem [{\citenamefont {Hirschorn}\ and\ \citenamefont
  {Orazem}(2009)}]{Hirschorn2009}%
  \BibitemOpen
  \bibfield  {author} {\bibinfo {author} {\bibfnamefont {B.}~\bibnamefont
  {Hirschorn}}\ and\ \bibinfo {author} {\bibfnamefont {M.~E.}\ \bibnamefont
  {Orazem}},\ }\href {https://doi.org/10.1149/1.3190160} {\bibfield  {journal}
  {\bibinfo  {journal} {J. Electrochem. Soc.}\ }\textbf {\bibinfo {volume}
  {156}},\ \bibinfo {pages} {345} (\bibinfo {year} {2009})}\BibitemShut
  {NoStop}%
\bibitem [{\citenamefont {Walls}\ and\ \citenamefont
  {Milburn}(2008)}]{Walls2008}%
  \BibitemOpen
  \bibfield  {author} {\bibinfo {author} {\bibfnamefont {D.~F.}\ \bibnamefont
  {Walls}}\ and\ \bibinfo {author} {\bibfnamefont {G.~J.}\ \bibnamefont
  {Milburn}},\ }\href {https://www.springer.com/gp/book/9783540285731} {\emph
  {\bibinfo {title} {{Quantum Optics}}}},\ \bibinfo {edition} {2nd}\ ed.\
  (\bibinfo  {publisher} {Springer},\ \bibinfo {year} {2008})\BibitemShut
  {NoStop}%
\bibitem [{\citenamefont {Isogai}\ \emph {et~al.}(2013)\citenamefont {Isogai},
  \citenamefont {Miller}, \citenamefont {Kwee}, \citenamefont {Barsotti},\ and\
  \citenamefont {Evans}}]{Isogai2013}%
  \BibitemOpen
  \bibfield  {author} {\bibinfo {author} {\bibfnamefont {T.}~\bibnamefont
  {Isogai}}, \bibinfo {author} {\bibfnamefont {J.}~\bibnamefont {Miller}},
  \bibinfo {author} {\bibfnamefont {P.}~\bibnamefont {Kwee}}, \bibinfo {author}
  {\bibfnamefont {L.}~\bibnamefont {Barsotti}}, \ and\ \bibinfo {author}
  {\bibfnamefont {M.}~\bibnamefont {Evans}},\ }\href
  {https://www.osapublishing.org/oe/abstract.cfm?uri=oe-21-24-30114} {\bibfield
   {journal} {\bibinfo  {journal} {Opt. Express}\ }\textbf {\bibinfo {volume}
  {21}},\ \bibinfo {pages} {30114} (\bibinfo {year} {2013})}\BibitemShut
  {NoStop}%
\bibitem [{\citenamefont {Oelker}\ \emph
  {et~al.}(2016{\natexlab{b}})\citenamefont {Oelker}, \citenamefont {Isogai},
  \citenamefont {Miller}, \citenamefont {Tse}, \citenamefont {Barsotti},
  \citenamefont {Mavalvala},\ and\ \citenamefont {Evans}}]{Oelker2016a}%
  \BibitemOpen
  \bibfield  {author} {\bibinfo {author} {\bibfnamefont {E.}~\bibnamefont
  {Oelker}}, \bibinfo {author} {\bibfnamefont {T.}~\bibnamefont {Isogai}},
  \bibinfo {author} {\bibfnamefont {J.}~\bibnamefont {Miller}}, \bibinfo
  {author} {\bibfnamefont {M.}~\bibnamefont {Tse}}, \bibinfo {author}
  {\bibfnamefont {L.}~\bibnamefont {Barsotti}}, \bibinfo {author}
  {\bibfnamefont {N.}~\bibnamefont {Mavalvala}}, \ and\ \bibinfo {author}
  {\bibfnamefont {M.}~\bibnamefont {Evans}},\ }\href
  {http://dx.doi.org/10.1103/PhysRevLett.116.041102} {\bibfield  {journal}
  {\bibinfo  {journal} {Phys. Rev. Lett.}\ }\textbf {\bibinfo {volume} {116}},\
  \bibinfo {pages} {041102} (\bibinfo {year} {2016}{\natexlab{b}})}\BibitemShut
  {NoStop}%
\bibitem [{\citenamefont {Bogan}\ \emph {et~al.}(2015)\citenamefont {Bogan},
  \citenamefont {Kwee}, \citenamefont {Hild}, \citenamefont {Huttner},\ and\
  \citenamefont {Willke}}]{Bogan2015}%
  \BibitemOpen
  \bibfield  {author} {\bibinfo {author} {\bibfnamefont {C.}~\bibnamefont
  {Bogan}}, \bibinfo {author} {\bibfnamefont {P.}~\bibnamefont {Kwee}},
  \bibinfo {author} {\bibfnamefont {S.}~\bibnamefont {Hild}}, \bibinfo {author}
  {\bibfnamefont {S.~H.}\ \bibnamefont {Huttner}}, \ and\ \bibinfo {author}
  {\bibfnamefont {B.}~\bibnamefont {Willke}},\ }\href
  {https://www.osapublishing.org/abstract.cfm?URI=oe-23-12-15380} {\bibfield
  {journal} {\bibinfo  {journal} {Opt. Express}\ }\textbf {\bibinfo {volume}
  {23}},\ \bibinfo {pages} {15380} (\bibinfo {year} {2015})}\BibitemShut
  {NoStop}%
\bibitem [{\citenamefont {Helstrom}(1967)}]{Helstrom1967}%
  \BibitemOpen
  \bibfield  {author} {\bibinfo {author} {\bibfnamefont {C.}~\bibnamefont
  {Helstrom}},\ }\href
  {https://www.sciencedirect.com/science/article/pii/0375960167903660?via%3Dihub}
  {\bibfield  {journal} {\bibinfo  {journal} {Physics Letters A}\ }\textbf
  {\bibinfo {volume} {25}},\ \bibinfo {pages} {101} (\bibinfo {year}
  {1967})}\BibitemShut {NoStop}%
\bibitem [{\citenamefont {Holevo}(2011)}]{Holevo2011}%
  \BibitemOpen
  \bibfield  {author} {\bibinfo {author} {\bibfnamefont {A.}~\bibnamefont
  {Holevo}},\ }\href {https://doi.org/10.1007/978-88-7642-378-9} {\emph
  {\bibinfo {title} {{Probabilistic and Statistical Aspects of quantum
  theory}}}},\ \bibinfo {edition} {2nd}\ ed.\ (\bibinfo  {publisher} {Scuola
  Normale Superiore},\ \bibinfo {year} {2011})\BibitemShut {NoStop}%
\bibitem [{\citenamefont {Braunstein}\ \emph {et~al.}(1996)\citenamefont
  {Braunstein}, \citenamefont {Caves},\ and\ \citenamefont
  {Milburn}}]{Braunstein1996}%
  \BibitemOpen
  \bibfield  {author} {\bibinfo {author} {\bibfnamefont {S.~L.}\ \bibnamefont
  {Braunstein}}, \bibinfo {author} {\bibfnamefont {C.~M.}\ \bibnamefont
  {Caves}}, \ and\ \bibinfo {author} {\bibfnamefont {G.~J.}\ \bibnamefont
  {Milburn}},\ }\href
  {http://linkinghub.elsevier.com/retrieve/pii/S0003491696900408} {\bibfield
  {journal} {\bibinfo  {journal} {Annals of Physics}\ }\textbf {\bibinfo
  {volume} {247}},\ \bibinfo {pages} {135} (\bibinfo {year}
  {1996})}\BibitemShut {NoStop}%
\bibitem [{\citenamefont {Giovannetti}\ \emph {et~al.}(2011)\citenamefont
  {Giovannetti}, \citenamefont {Lloyd},\ and\ \citenamefont
  {Maccone}}]{Giovannetti2011}%
  \BibitemOpen
  \bibfield  {author} {\bibinfo {author} {\bibfnamefont {V.}~\bibnamefont
  {Giovannetti}}, \bibinfo {author} {\bibfnamefont {S.}~\bibnamefont {Lloyd}},
  \ and\ \bibinfo {author} {\bibfnamefont {L.}~\bibnamefont {Maccone}},\ }\href
  {http://www.nature.com/doifinder/10.1038/nphoton.2011.35} {\bibfield
  {journal} {\bibinfo  {journal} {Nat. Photonics}\ }\textbf {\bibinfo {volume}
  {5}},\ \bibinfo {pages} {222} (\bibinfo {year} {2011})}\BibitemShut {NoStop}%
\bibitem [{\citenamefont {Tsang}\ \emph {et~al.}(2011)\citenamefont {Tsang},
  \citenamefont {Wiseman},\ and\ \citenamefont {Caves}}]{Tsang2011}%
  \BibitemOpen
  \bibfield  {author} {\bibinfo {author} {\bibfnamefont {M.}~\bibnamefont
  {Tsang}}, \bibinfo {author} {\bibfnamefont {H.~M.}\ \bibnamefont {Wiseman}},
  \ and\ \bibinfo {author} {\bibfnamefont {C.~M.}\ \bibnamefont {Caves}},\
  }\href {http://journals.aps.org/prl/abstract/10.1103/PhysRevLett.106.090401}
  {\bibfield  {journal} {\bibinfo  {journal} {Phys. Rev. Lett.}\ }\textbf
  {\bibinfo {volume} {106}},\ \bibinfo {pages} {90401} (\bibinfo {year}
  {2011})}\BibitemShut {NoStop}%
\bibitem [{\citenamefont {Miao}\ \emph {et~al.}(2017)\citenamefont {Miao},
  \citenamefont {Adhikari}, \citenamefont {Ma}, \citenamefont {Pang},\ and\
  \citenamefont {Chen}}]{Miaoa}%
  \BibitemOpen
  \bibfield  {author} {\bibinfo {author} {\bibfnamefont {H.}~\bibnamefont
  {Miao}}, \bibinfo {author} {\bibfnamefont {R.~X.}\ \bibnamefont {Adhikari}},
  \bibinfo {author} {\bibfnamefont {Y.}~\bibnamefont {Ma}}, \bibinfo {author}
  {\bibfnamefont {B.}~\bibnamefont {Pang}}, \ and\ \bibinfo {author}
  {\bibfnamefont {Y.}~\bibnamefont {Chen}},\ }\href
  {https://arxiv.org/pdf/1608.00766.pdf} {\bibfield  {journal} {\bibinfo
  {journal} {Phys. Rev. Lett.}\ }\textbf {\bibinfo {volume} {119}},\ \bibinfo
  {pages} {050801} (\bibinfo {year} {2017})}\BibitemShut {NoStop}%
\bibitem [{\citenamefont {Caves}\ and\ \citenamefont
  {Schumaker}(1985)}]{Caves1985}%
  \BibitemOpen
  \bibfield  {author} {\bibinfo {author} {\bibfnamefont {C.~M.}\ \bibnamefont
  {Caves}}\ and\ \bibinfo {author} {\bibfnamefont {B.~L.}\ \bibnamefont
  {Schumaker}},\ }\href {https://doi.org/10.1103/PhysRevA.31.3068} {\bibfield
  {journal} {\bibinfo  {journal} {Phys. Rev. A}\ }\textbf {\bibinfo {volume}
  {31}},\ \bibinfo {pages} {3068} (\bibinfo {year} {1985})}\BibitemShut
  {NoStop}%
\bibitem [{\citenamefont {Schumaker}\ and\ \citenamefont
  {Caves}(1985)}]{Schumaker1985}%
  \BibitemOpen
  \bibfield  {author} {\bibinfo {author} {\bibfnamefont {B.~L.}\ \bibnamefont
  {Schumaker}}\ and\ \bibinfo {author} {\bibfnamefont {C.~M.}\ \bibnamefont
  {Caves}},\ }\href {https://doi.org/10.1103/PhysRevA.31.3093} {\bibfield
  {journal} {\bibinfo  {journal} {Phys. Rev. A}\ }\textbf {\bibinfo {volume}
  {31}},\ \bibinfo {pages} {3093} (\bibinfo {year} {1985})}\BibitemShut
  {NoStop}%
\bibitem [{\citenamefont {C{\'{e}}sar}\ \emph {et~al.}(2009)\citenamefont
  {C{\'{e}}sar}, \citenamefont {Coelho}, \citenamefont {Cassemiro},
  \citenamefont {Villar}, \citenamefont {Lassen}, \citenamefont {Nussenzveig},\
  and\ \citenamefont {Martinelli}}]{Cesar2009}%
  \BibitemOpen
  \bibfield  {author} {\bibinfo {author} {\bibfnamefont {J.~E.}\ \bibnamefont
  {C{\'{e}}sar}}, \bibinfo {author} {\bibfnamefont {A.~S.}\ \bibnamefont
  {Coelho}}, \bibinfo {author} {\bibfnamefont {K.~N.}\ \bibnamefont
  {Cassemiro}}, \bibinfo {author} {\bibfnamefont {A.~S.}\ \bibnamefont
  {Villar}}, \bibinfo {author} {\bibfnamefont {M.}~\bibnamefont {Lassen}},
  \bibinfo {author} {\bibfnamefont {P.}~\bibnamefont {Nussenzveig}}, \ and\
  \bibinfo {author} {\bibfnamefont {M.}~\bibnamefont {Martinelli}},\ }\href
  {https://doi.org/10.1103/PhysRevA.79.063816} {\bibfield  {journal} {\bibinfo
  {journal} {Phys. Rev. A}\ }\textbf {\bibinfo {volume} {79}},\ \bibinfo
  {pages} {063816} (\bibinfo {year} {2009})}\BibitemShut {NoStop}%
\bibitem [{\citenamefont {Bechhoefer}(2005)}]{Bechhoefer2005}%
  \BibitemOpen
  \bibfield  {author} {\bibinfo {author} {\bibfnamefont {J.}~\bibnamefont
  {Bechhoefer}},\ }\href {https://link.aps.org/doi/10.1103/RevModPhys.77.783}
  {\bibfield  {journal} {\bibinfo  {journal} {Rev. Mod. Phys.}\ }\textbf
  {\bibinfo {volume} {77}},\ \bibinfo {pages} {783} (\bibinfo {year}
  {2005})}\BibitemShut {NoStop}%
\end{thebibliography}%

\end{document}